\documentclass[twocolumn,linenumbers]{aastex631}

\hypersetup{linkcolor=magenta,citecolor=blue,filecolor=cyan,urlcolor=purple}

\usepackage{graphicx}
\usepackage[caption=false]{subfig}
\usepackage{amsmath}

\usepackage{array,booktabs,threeparttable}
\usepackage{tabularx}

\received{2026 June 11}
\revised{2026 August 31}
\accepted{2026 September 9}

\shorttitle{The 3D Structure of CMEs in the Corona}
\shortauthors{Palmerio et al.}

\graphicspath{{./}{figures/}}



\begin{document}

\title{On the 3D Magnetic Structure of Coronal Mass Ejections Through the Solar Corona}

\correspondingauthor{Erika Palmerio}
\email{epalmerio@predsci.com}

\author[0000-0001-6590-3479]{Erika Palmerio}
\affiliation{Predictive Science Inc., San Diego, CA 92121, USA}

\author[0000-0002-9164-0008]{Michal Ben-Nun}
\affiliation{Predictive Science Inc., San Diego, CA 92121, USA}

\author[0000-0003-3843-3242]{Tibor T{\"o}r{\"o}k}
\affiliation{Predictive Science Inc., San Diego, CA 92121, USA}

\author[0000-0003-1759-4354]{Cooper Downs}
\affiliation{Predictive Science Inc., San Diego, CA 92121, USA}

\author[0000-0001-7053-4081]{Viacheslav S.\ Titov}
\affiliation{Predictive Science Inc., San Diego, CA 92121, USA}


\begin{abstract}

The magnetic fields that make up the internal structure of coronal mass ejections (CMEs) are thought to be organised in a flux rope configuration, consisting of twisted magnetic fields that wind about a central axis. Remote-sensing observations of CMEs show a wide range of morphologies, dynamics, and evolution, including rotation, non-uniform expansion, deflection, and interaction with the ambient solar wind. In-situ measurements, however, typically consist of a single 1D spacecraft trajectory through a large 3D structure (or few at best), restricting our ability to determine how CME magnetic structure varies in space and time. In this work, we analyse the magnetic configuration of two idealised CMEs during their early evolution in the range 1--30\,$R_{\odot}$ using the Magnetohydrodynamic Algorithm Outside a Sphere code. The initial flux ropes \edit1{differ only in their orientation with respect to the global dipole and}{} erupt from a bipolar active region below the streamer belt in a simplified coronal environment. We deploy a fleet of synthetic spacecraft throughout the CMEs' paths at different combinations of latitudes, longitudes, and heliocentric distances. We identify and examine flux rope signatures in the synthetic in-situ profiles to characterise radial variations as well as latitudinal/longitudinal ones. We find that, even in a simplified CME erupting under solar minimum-like conditions, the sampling location significantly affects the global structure that would be deduced from common flux rope reconstruction and analysis techniques used for in-situ measurements. Additionally, we consider possible in-situ signatures that may serve as indicators of the source region's topology.

\end{abstract}



\section{Introduction} \label{sec:intro}

Coronal mass ejections \citep[CMEs; e.g.,][]{webb2012} are humongous clouds of plasma and magnetic fields that are regularly ejected from the Sun into the heliosphere. Regardless of their pre-eruptive magnetic configuration \citep[e.g.,][]{patsourakos2020}, CMEs are believed to leave the Sun as flux ropes, i.e.\ helical structures that consist of twisted magnetic fields winding about a central axis \citep[e.g.,][]{green2018}. The paradigm of ``CMEs as flux ropes'' has been built over decades of research based on remote-sensing observations of the lower solar atmosphere \citep[e.g.,][]{rust1994, green2007} and the extended corona \citep[e.g.,][]{illing1985, cremades2004}, as well as in-situ measurements throughout the inner heliosphere \citep[e.g.,][]{burlaga1981, richardson2004c}. Flux rope models have been used extensively throughout the development of numerical simulations of CME eruptions, ranging from relatively idealised configurations \citep[e.g.,][]{roussev2003, fan2004} to more complex data-constrained formulations \citep[e.g.,][]{lynch2016b, torok2018}.

Despite continuous progress in both space instrumentation and computational capabilities, reconciling CME theories with observations remains an issue with several open questions even to this day. For example, CMEs appear in coronagraph data as 2D projections of large 3D structures, and their relation to flux ropes has to often be inferred with the aid of models and assumptions \citep[e.g.,][]{thernisien2006, howard2017}. Furthermore, in-situ measurements of CMEs only provide a single 1D trajectory---or a few, in the case of fortuitous multi-point encounters---through a much larger 3D cloud, and it is difficult to ascertain how representative such a sample is of the global flux rope configuration \citep[e.g.,][]{farrugia2011, mostl2012}. Even more so, the myriad analytical flux rope models that are commonly used to fit in-situ observations are each based on their own set of free parameters and constraints \citep[e.g.,][]{lepping1990, vandas1997, hu2002, nieveschinchilla2018b, weiss2024}, and it has been shown that it is not unusual for different models to retrieve significantly different CME configurations \citep[e.g.,][]{riley2004, alhaddad2018}.

In the in-situ regime, increasing interest has been given to multi-point measurements and their usefulness for reaching a deeper understanding of the magnetic structure and evolution of CMEs \citep[e.g.,][]{palmerio2021d, palmerio2025, davies2021, davies2022, mostl2022}, with particular attention on radial \citep[e.g.,][]{good2018, vrsnak2019}, longitudinal \citep[e.g.,][]{lugaz2024, banu2025}, and mesoscale \citep[e.g.,][]{lugaz2018, palmerio2024} variations. However, all multi-point studies performed to date have relied on serendipitous spacecraft alignments and not on coordinated efforts characterised by well-defined separations in latitude, longitude, and heliocentric distance. In the absence of dedicated multi-spacecraft missions that allow sampling of the internal structure of CMEs in a systematic manner, the issue has been tackled in the existing literature via analytical modelling and magnetohydrodynamic (MHD) simulations.

Studies that have analysed multiple CME in-situ profiles in computational settings involve the placement of so-called synthetic spacecraft within the computational domain. These virtual probes can be, e.g., radially distributed along the Sun--observer line to explore how CMEs evolve as they propagate away from the Sun \citep[e.g.,][]{alhaddad2019b, scolini2021a, verbeke2022, regnault2023}. Another approach is to place a swarm of synthetic spacecraft uniformly distributed in latitude and longitude around the CME propagation direction and over different heliocentric distances, as was done by \citet{scolini2021b, scolini2023} to examine variations in CME complexity and---in a simplified, analytical description---by \citet{rudisser2024} to explore the effect of spacecraft trajectory on the detected in-situ profiles. \edit1{Alternatively, several in-situ profiles can be generated and analysed along realistic trajectories of proposed multi-spacecraft mission concepts, as demonstrated by \citet{manchester2025}.} Most of these works have focussed on the interplanetary evolution of CMEs (with a simulation domain extending up to 1~au and beyond), but some analyses have been dedicated to the magnetic structure of CMEs in the solar corona. For example, \citet{alhaddad2019a} analysed two simulated CMEs (a `twisted' and a `writhed' one) by placing a grid of virtual spacecraft at 15\,$R_{\odot}$, and concluded that flux rope models cannot reliably distinguish between the two. Additionally, \citet{lynch2022} studied the applicability of flux rope fitting techniques in the range 10--30\,$R_{\odot}$ using a set of stationary synthetic probes as well as a set of moving observers based on actual ephemeris data from the Parker Solar Probe \citep{fox2016} spacecraft.

In this work, we perform a detailed analysis of the 3D magnetic structure of CMEs through the extended solar corona. We explore two separate MHD simulations in which the involved flux ropes are erupted in a simplified coronal configuration, i.e.\ from a bipolar active region located under the streamer belt---the two runs differing uniquely by the orientation of the source region polarities. The two CMEs propagate through a uniform solar wind background, thus reducing to a minimum effects arising from interactions with the ambient medium, which are well known to lead to significant changes to CME direction, morphology, and internal structure \citep[e.g.,][]{winslow2021, palmerio2022b}. We place a fleet of synthetic spacecraft at different combinations of latitudes, longitudes, and heliocentric distances in both simulations, in order to identify and examine flux rope signatures in the resulting in-situ profiles. Our aim is to characterise radial variations as well as latitudinal/longitudinal ones during the early evolution of CMEs using analysis tools and reconstruction techniques that are commonly applied to actual in-situ measurements. We expect our findings to aid analysis and interpretation of the internal structure of CMEs observed by Parker Solar Probe below 30\,$R_{\odot}$ \citep[e.g.,][]{romeo2023}. The article is organised as follows. In Section~\ref{sec:setup}, we provide an overview of the MHD simulation setup and describe the two CME runs. In Section~\ref{sec:results} we present our resulting synthetic in-situ profiles, which are analysed and interpreted in Section~\ref{sec:interpretation} and further discussed in Section~\ref{sec:discussion}. Finally, in Section~\ref{sec:conclusions} we summarise our findings and draw our conclusions.


\section{Simulation Setup} \label{sec:setup}

In this work, we analyse in detail two CME simulations reported in \citet{bennun2023}. They are performed using the Magnetohydrodynamic Algorithm Outside a Sphere \citep[MAS; e.g.,][]{mikic1999, mikic2018, downs2025} code, which solves the time-dependent resistive MHD equations on a non-uniform spherical mesh. In this application, MAS is run in the so-called polytropic approximation \citep[e.g.,][]{linker1999, linker2003, lionello2013}, which employs an adiabatic energy equation with a small, fixed polytropic index---set, in this case, to $\gamma = 1.05$---that captures the high conductivity of low-beta coronal plasmas without solving for the more expensive thermodynamic terms. The background configuration of the corona is modelled to represent a simplified yet nearly-realistic environment, i.e.\ consisting of a global dipole field that leads (upon MHD relaxation) to the formation of a streamer belt and a heliospheric current sheet, resembling solar minimum conditions. Additionally, these simulations employ a simplified solar wind, namely the roughly-uniform slow wind that results naturally from a polytropic model, thus neglecting additional waves and heating that are able to accelerate fast streams. This setup limits CME interactions with a structured background environment, thus allowing us to focus on the intrinsic structure and evolution of solar eruptions as they propagate through the corona. The full spatial computational domain covers the range 1--30\,$R_{\odot}$ in the radial direction ($r$), ${\pm}90^{\circ}$ in the latitudinal direction ($\theta$), and ${\pm}180^{\circ}$ in the longitudinal direction ($\phi$). \edit1{The domain is discretised on a nonuniform mesh of $405\times368\times318$ points in ($r$, $\theta$, $\phi$), with the radial resolution varying from $0.002\,R_{\odot}$ close to the photosphere to $0.413\,R_{\odot}$ at the outer boundary, whilst the angular resolutions are refined around the CME source region and progressively coarsened away from it.}

\begin{figure*}[th!]
\centering
\includegraphics[width=0.495\linewidth]{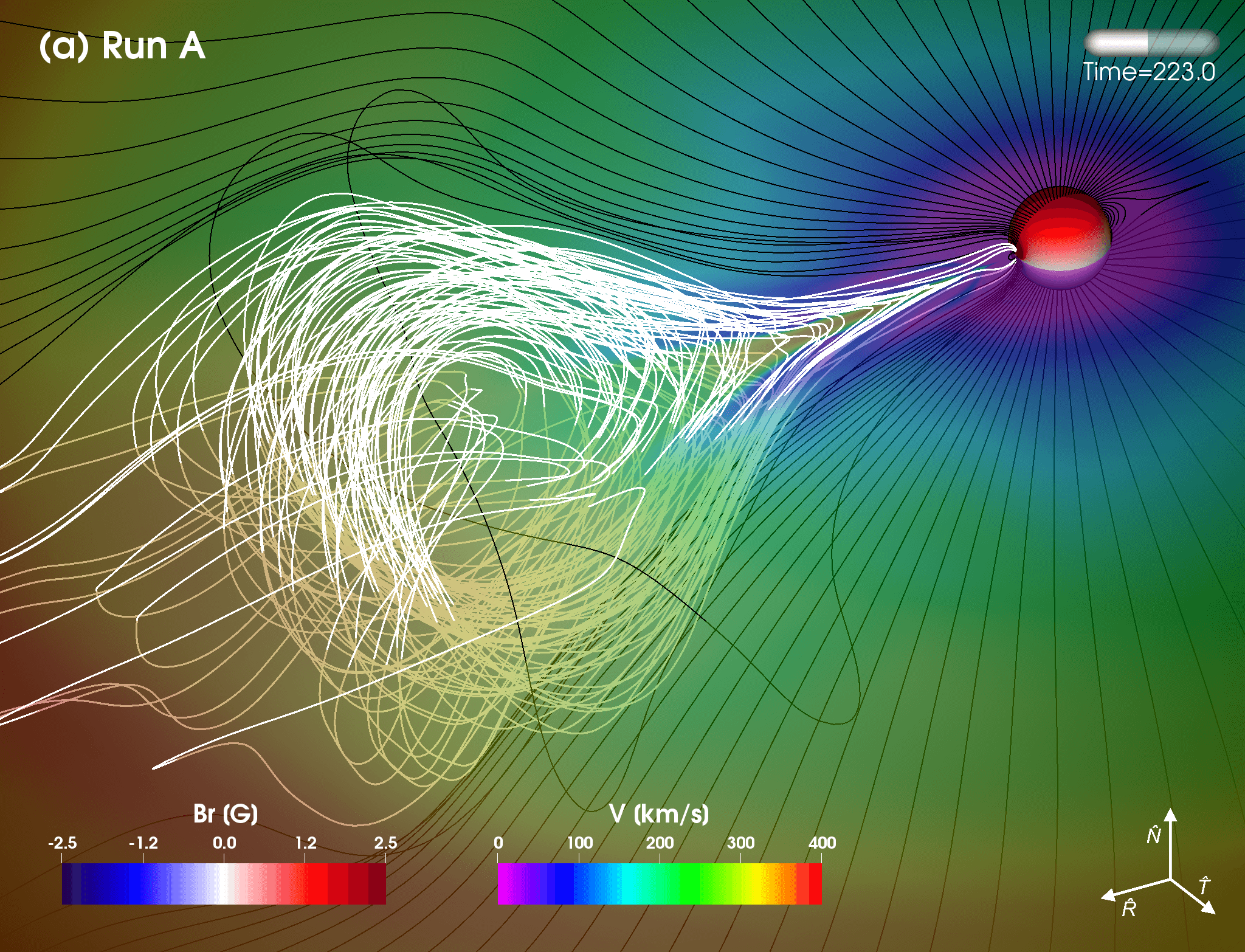}
\includegraphics[width=0.495\linewidth]{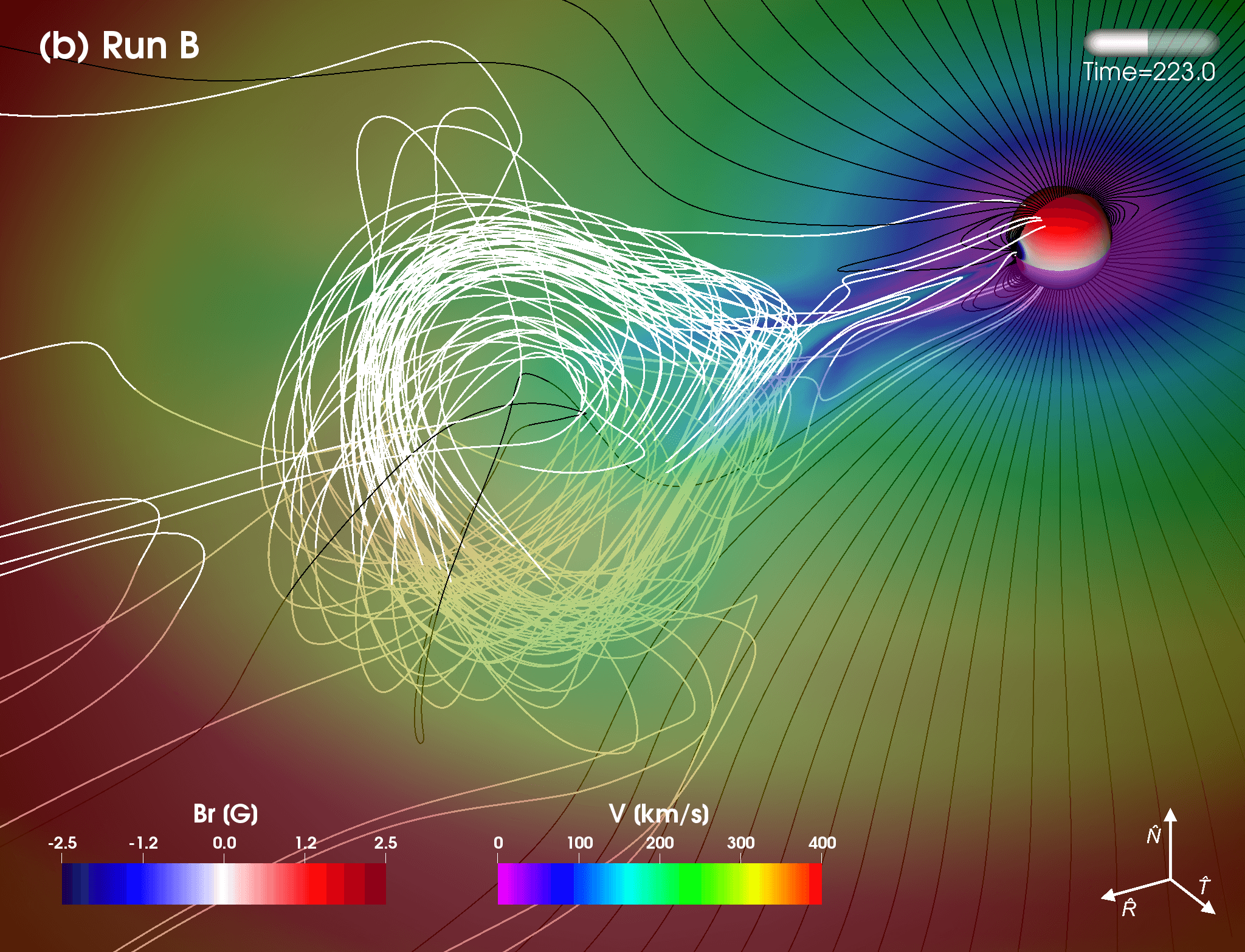}\\
\vspace*{.03in}
\includegraphics[width=0.495\linewidth]{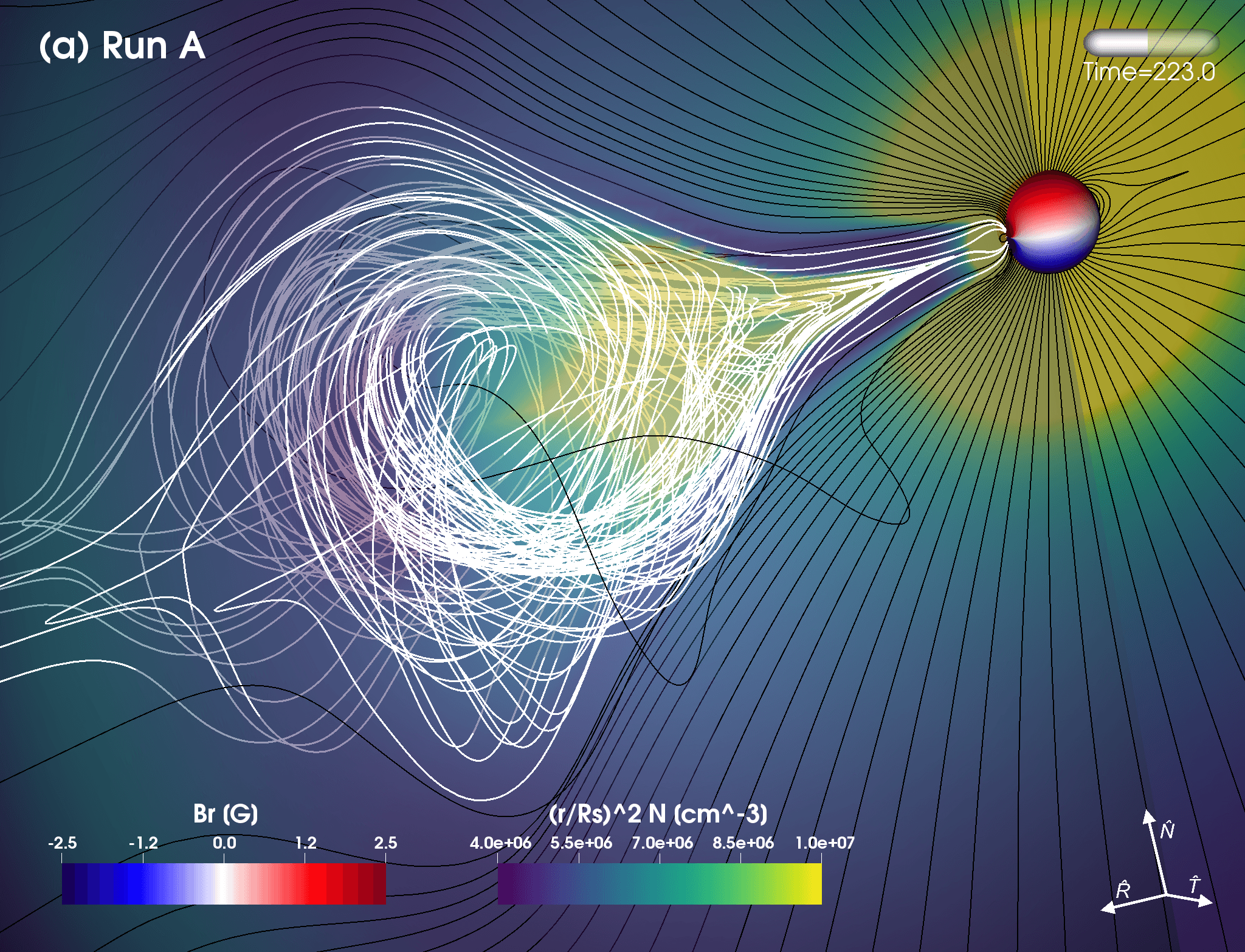}
\includegraphics[width=0.495\linewidth]{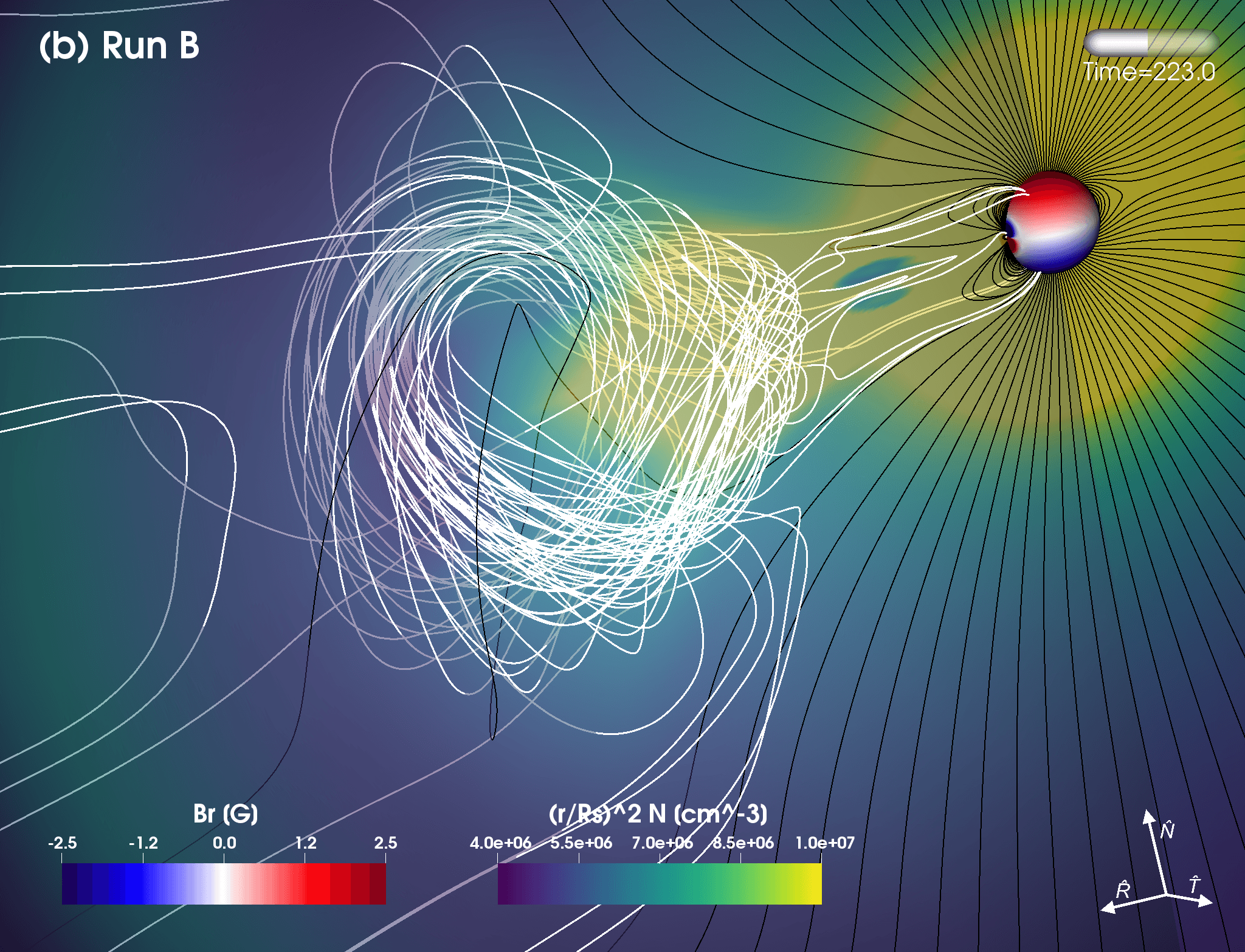}
\caption{Overview of the two simulated CMEs investigated in this study: (a) Run~A and (b) Run~B. In all panels, representative magnetic field lines are shown, coloured to approximately distinguish between the (black) ambient coronal magnetic field and the (white) CME flux rope. The top panels show the field lines overlaid on the equatorial solar wind speed, while the bottom panels show them together with the (scaled) coronal density in the meridional plane containing the fictitious Sun--Earth line.
\label{fig:cme}}
\end{figure*}

\begin{figure*}[th!]
\centering
\includegraphics[width=0.495\linewidth]{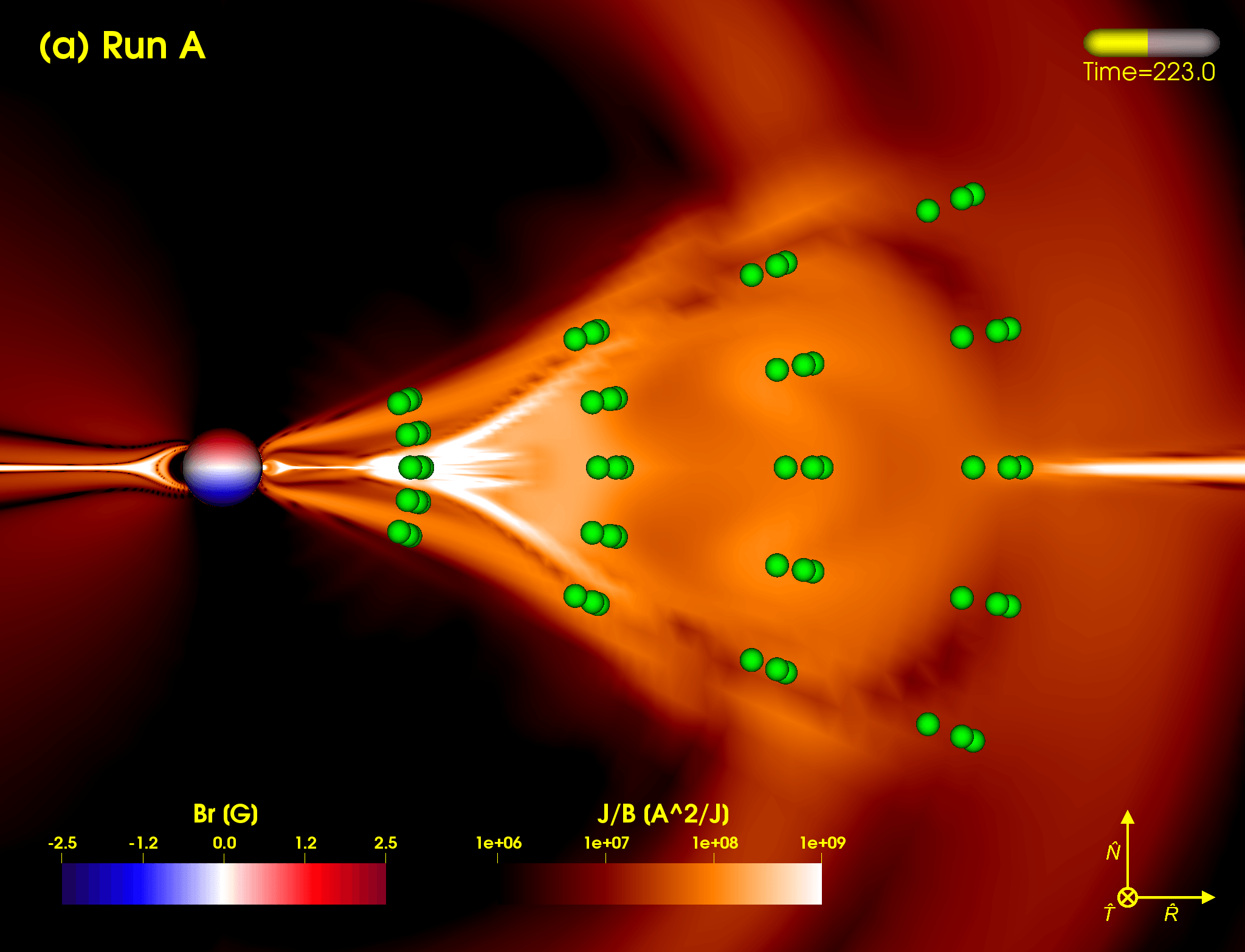}
\includegraphics[width=0.495\linewidth]{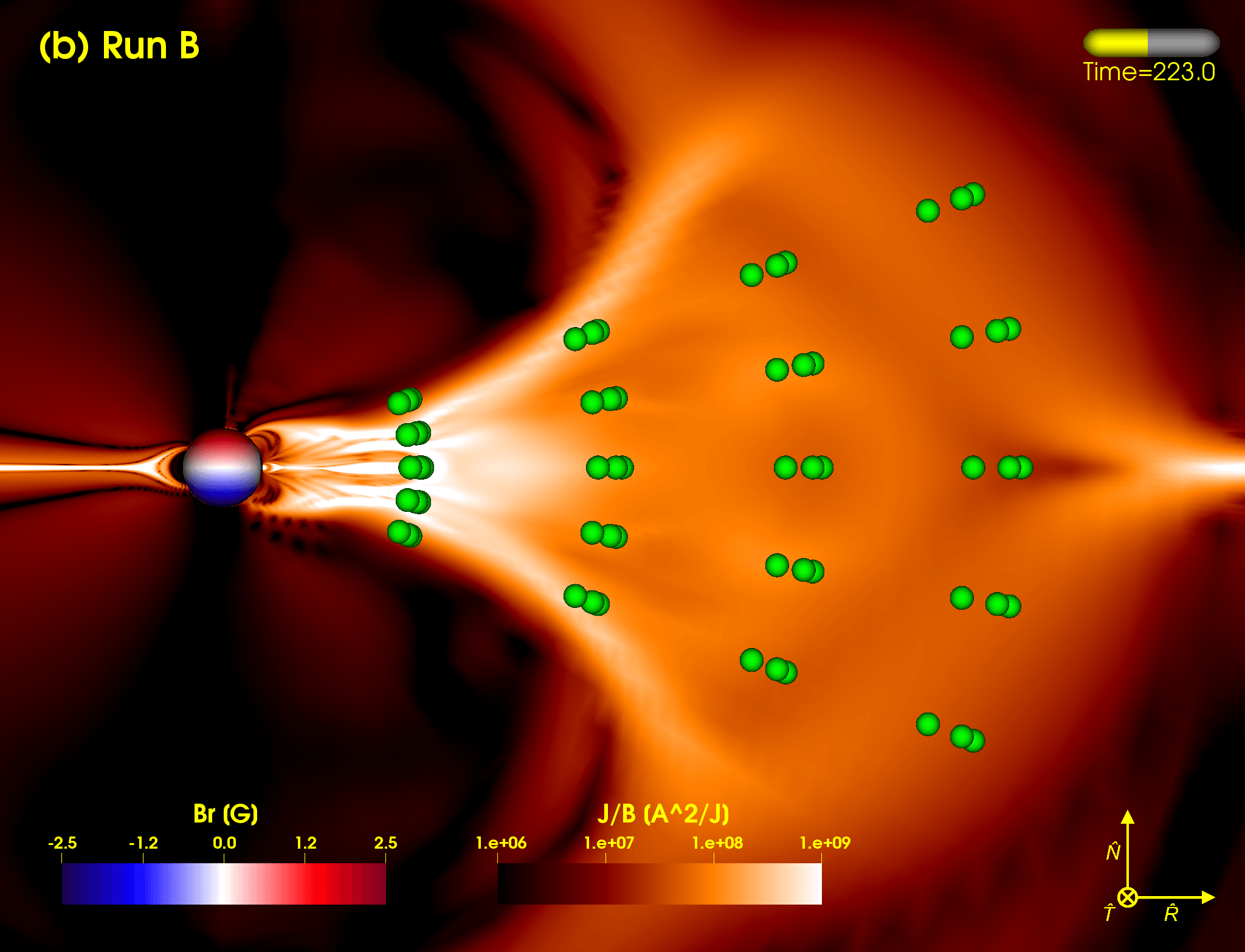}
\caption{Spatial distribution of the synthetic spacecraft for (a) Run~A and (b) Run~B projected onto the meridional plane at $\phi = 0^{\circ}$. The virtual probes are represented by green spheres. The meridional plane is shown by means of the normalised current density $|J/B|$ at $t \simeq 9.5$~h, i.e.\ when the CME apex reaches $r \simeq 20\,R_{\odot}$ in either run. The animated version of this figure shows the evolution of $|J/B|$ throughout the temporal domain of the simulations (0--20~h).\\
(An animation of this figure is available.)
\label{fig:sc}}
\end{figure*}

The full set of simulations in \citet{bennun2023} consists of eight model runs, realised via different combinations of CME source region location (positioned either under the streamer belt, at $\theta=0^{\circ}$, or on the northern hemisphere, at $\theta=45^{\circ}$), magnetic orientation of the source region (either north--south or south--north), and flux rope chirality (either right- or left-handed). In this work, we focus on the two right-handed runs that originate from below the streamer belt---specifically, Run~1 and Run~3 in \citet{bennun2023}, \edit1{hereafter referred to as Run~A and Run~B, respectively, and presented in Figure~\ref{fig:cme}}. The CME source regions are modelled as bipolar active regions that contain flux ropes with an axial current density slightly above (${\sim}1$\%) their stable-equilibrium configuration, so that an eruption can be triggered. To achieve this, we use the modified Titov--D{\'e}moulin \citep[TDm;][]{titov2014} flux rope model, which is an extension of the Titov--D{\'e}moulin \citep[TD;][]{titov1999} one. In Run~A \citep[Run~1 in][]{bennun2023} the modelled active region has the same magnetic orientation (north--south) as the global solar dipole, resulting in a bipolar CME source region, whilst in Run~B \citep[Run~3 in][]{bennun2023} the modelled active region has opposite magnetic orientation (south--north) to the global solar dipole, resulting in a quadrupolar CME source region. As shown in \citet{bennun2023}, both runs result in the CMEs propagating radially along the streamer belt, i.e., they do not exhibit any signature of deflection (see e.g.\ their Figure~8 \edit1{as well as Figure~\ref{fig:cme} in this work}).

To investigate the 3D magnetic structure and evolution of the two modelled eruptions as they travel away from the Sun, we place a swarm of synthetic spacecraft around the propagation direction of the CMEs. If we assume the source region(s) to be located at $(\theta, \phi) = (0^{\circ}, 0^{\circ})$---ideally representing the Sun--Earth line---the virtual probes occupy the range $r = [5, 20]\,R_{\odot}$ in increments of $5\,R_{\odot}$, $\theta = [-20^{\circ}, 20^{\circ}]$ in increments of $10^{\circ}$, and $\phi = [-20^{\circ}, 20^{\circ}]$ in increments of $10^{\circ}$. This results in a total of 100 synthetic spacecraft (per simulation run) that realise different sampling trajectories through the CMEs---at the nose, at intermediate distances, and at the flanks---as shown in Figure~\ref{fig:sc}. We remark that the observers are set to be fixed in the corotating frame, as to avoid additional complexities due to a specific spacecraft's time-dependent trajectory. For each virtual probe, we extract the time-dependent MHD quantities for magnetic field and plasma as it would be done by actual spacecraft monitoring the in-situ conditions in interplanetary space. To emulate measurements produced by real heliophysics missions, we convert the magnetic field data from simulation-specific coordinates into the \edit1{heliocentric} radial--tangential--normal (RTN) frame. The extraction of  magnetic field and plasma quantities is performed for both runs throughout the temporal computational domain, i.e.\ 0--20~h, \edit1{at a cadence of 12~minutes}.


\section{Synthetic In-situ Profiles} \label{sec:results}

Once the synthetic spacecraft have been placed and the different in-situ time series have been obtained (100 per simulation run), we examine the resulting profiles to characterise the two CMEs from both a structural (i.e., concerning latitudinal/longitudinal variations) and an evolutionary (i.e., relating to radial changes) standpoint. In this section, we focus on the most noteworthy results emerging from such analysis, starting with the examination of the in-situ profiles along the CME propagation direction, where---at least intuitively---it is expected to observe the clearest, least complex signatures.

\subsection{Radial Evolution Along the CME Nose Path} \label{subsec:nose}

\begin{figure*}[p!]
\centering
\includegraphics[width=0.99\linewidth]{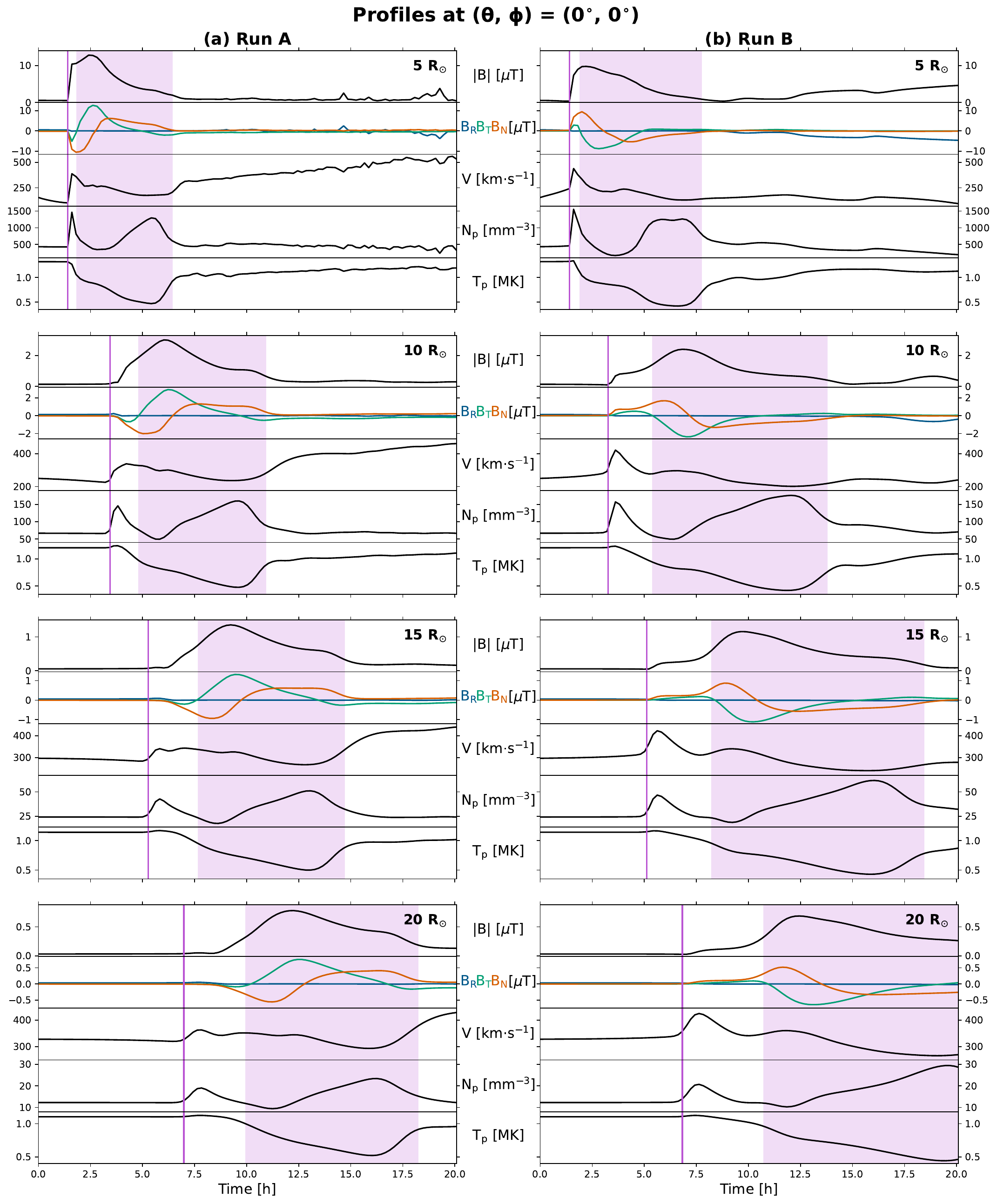}
\caption{Synthetic spacecraft profiles for fixed $(\theta, \phi) = (0^{\circ}, 0^{\circ})$ at four radial distances (5, 10, 15, 20\,$R_{\odot}$) for (a) Run~A and (b) Run~B. Each plot shows, from top to bottom: magnetic field magnitude, magnetic field components in RTN coordinates, solar wind speed, density, and temperature. CME-driven sheaths are marked by vertical lines, whilst CME flux rope intervals are indicated by shaded areas. The $y$-axes are scaled uniformly across each row to highlight differences between the two runs. \label{fig:cmenose} }
\end{figure*}

The first step of our analysis of in-situ profiles is to inspect the magnetic field and plasma time series along the CME propagation direction (or CME nose), i.e.\ at $(\theta, \phi) = (0^{\circ}, 0^{\circ})$. These profiles are shown in Figure~\ref{fig:cmenose}, allowing us to evaluate the presence of flux rope signatures and to visually determine the corresponding flux rope type(s)---evaluated from the sequence of magnetic field directions encountered during the spacecraft crossing (e.g., NES, NWS, ESW, etc.), where the three letters denote the \edit1{cardinal directions} of the magnetic field at the leading edge, centre, and trailing edge of the flux rope, respectively \citep[e.g.,][]{bothmer1998, mulligan1998, palmerio2018}. To identify the two structures in the MHD time series, we search for well-known in-situ CME signatures \citep[e.g.,][]{zurbuchen2006, kilpua2017b}, such as enhanced magnetic field magnitude, smoothly-rotating magnetic field vectors, declining velocity profile (signature of expansion), and depressed proton temperature. A smooth rotation of the magnetic field vectors is characteristic of so-called magnetic clouds \citep{burlaga1981}, or more generally of ejecta that are interpreted to display a flux rope configuration. \edit1{We identify the boundaries and flux rope types of the two CMEs through visual inspection of the in-situ time series, following the same approach commonly applied to real spacecraft observations.}

Figure~\ref{fig:cmenose} shows, marked in violet, the passage of the CME-driven sheath as well as the extent of the CME ejecta at each radial distance (5, 10, 15, and 20\,$R_{\odot}$) for both runs. We remark that, here, we use the more general terminology of ``CME-driven sheath'' rather than ``CME-driven shock'' since a clear, simultaneous jump in magnetic field, speed, density, and temperature \citep[e.g.,][]{kilpua2015} is not always present in the observed time series. On the other hand, it is well known that CME-driven sheath regions can feature diverse properties, including the presence or lack thereof of a preceding shock \citep[e.g.,][]{salman2020b}. \edit1{It is also worth noting that the low adiabatic index ($\gamma = 1.05$, see Section~\ref{sec:setup}) adopted in our simulations substantially reduces the temperature jump across the CME-driven disturbance and, consequently, may weaken the corresponding sheath signatures.} In the two runs investigated here, a sharp and prominent enhancement in magnetic field, solar wind speed, and density is observed only at 5\,$R_{\odot}$. \edit1{This may result from the increasing ambient solar wind speed with heliocentric distance (from below 250~km$\cdot$s$^{-1}$ at 5\,$R_{\odot}$ to approximately 325~km$\cdot$s$^{-1}$ at 20\,$R_{\odot}$), which reduces the relative speed of a CME travelling at ${\sim}350$~km$\cdot$s$^{-1}$, and/or from the progressively coarser radial grid (with resolutions of 0.047, 0.122, 0.198, and 0.271\,$R_{\odot}$ at 5, 10, 15, and 20\,$R_{\odot}$, respectively), which may reduce the ability of the MHD scheme to capture the shock.} We also note that the CME ejecta in Run~A is followed by a period of enhanced solar wind speed, whereas this is not the case for Run~B. This difference is due to post-eruption reconnection flows \citep[e.g.,][]{reeves2015, slemzin2022}, since the modelled ambient corona considered in this work includes only a slow solar wind. This may also explain why the CME ejecta identified in Run~B systematically exhibits a longer duration than in Run~A: the presence of a trailing faster flow in the bipolar case leads to enhanced compression of the rear portion of the ejecta, as is evident in the speed profiles shown in Figure~\ref{fig:cmenose}.

Focussing now on the magnetic structure of the CME ejecta, the first notable difference is that the Run~B profiles are characterised by systematically lower magnetic field magnitudes than those in Run~A. This may result from the stronger expansion of the ejecta discussed above and/or from intrinsic evolution of the CME magnetic bundle through reconnection and erosion processes \citep[e.g.,][\edit1{see also Section~\ref{subsec:topology} for further discussion}]{lynch2013, hosteaux2021}. Despite this discrepancy, both runs clearly exhibit smoothly rotating magnetic field vectors, a characteristic signature of flux ropes. For Run~A, the magnetic field rotates from south to north ($B_\mathrm{N}$ component) whilst pointing towards the west at its centre ($B_\mathrm{T}$ component); for Run~B, the magnetic field rotates from north to south ($B_\mathrm{N}$ component) whilst pointing towards the east at its centre ($B_\mathrm{T}$ component). Hence, the CME in Run~A is of type south--west--north (SWN), whilst the CME in Run~B is of type north--east--south (NES). This classification is in agreement with the magnetic configuration of the corresponding flux ropes at the Sun \citep[cf.\ Figure~3 in][]{bennun2023}, i.e.\ of two right-handed structures with oppositely-directed axes that feature low inclinations to the solar equatorial plane. We note that both flux ropes maintain their magnetic structure (i.e., their flux rope type) across all the examined radial distances, as would be expected of CMEs that do not experience significant rotations, deflections, or interactions with a structured ambient wind. Additionally, the $B_\mathrm{R}$ component of the magnetic field displays near-zero values throughout the examined flux rope intervals, usually considered a signature of a spacecraft encounter close to both the CME nose and central axis \citep{burlaga1988}. 

For both runs, the only striking difference across the profiles with increasing heliocentric distance is that the flux rope ejecta exhibit progressively smoother and elongated magnetic field profiles, which can be interpreted as a natural consequence of radial expansion \citep[e.g.,][]{demoulin2009a}. As mentioned earlier, the ``central encounter profiles'' presented in this section are intuitively assumed to represent the least complex cases of spacecraft trajectories through a CME. Hence, we shall consider the flux rope structures investigated here as our ``baseline'' and explore, in the following sections, variations to the magnetic configuration emerging at varying latitudes and/or longitudes---i.e., at intermediate and flank encounters of our synthetic probes through the two simulated CMEs.

\subsection{Longitudinal Variations at 10\,$R_{\odot}$} \label{subsec:lonvar}

\begin{figure*}[p!]
\centering
\includegraphics[width=0.99\linewidth]{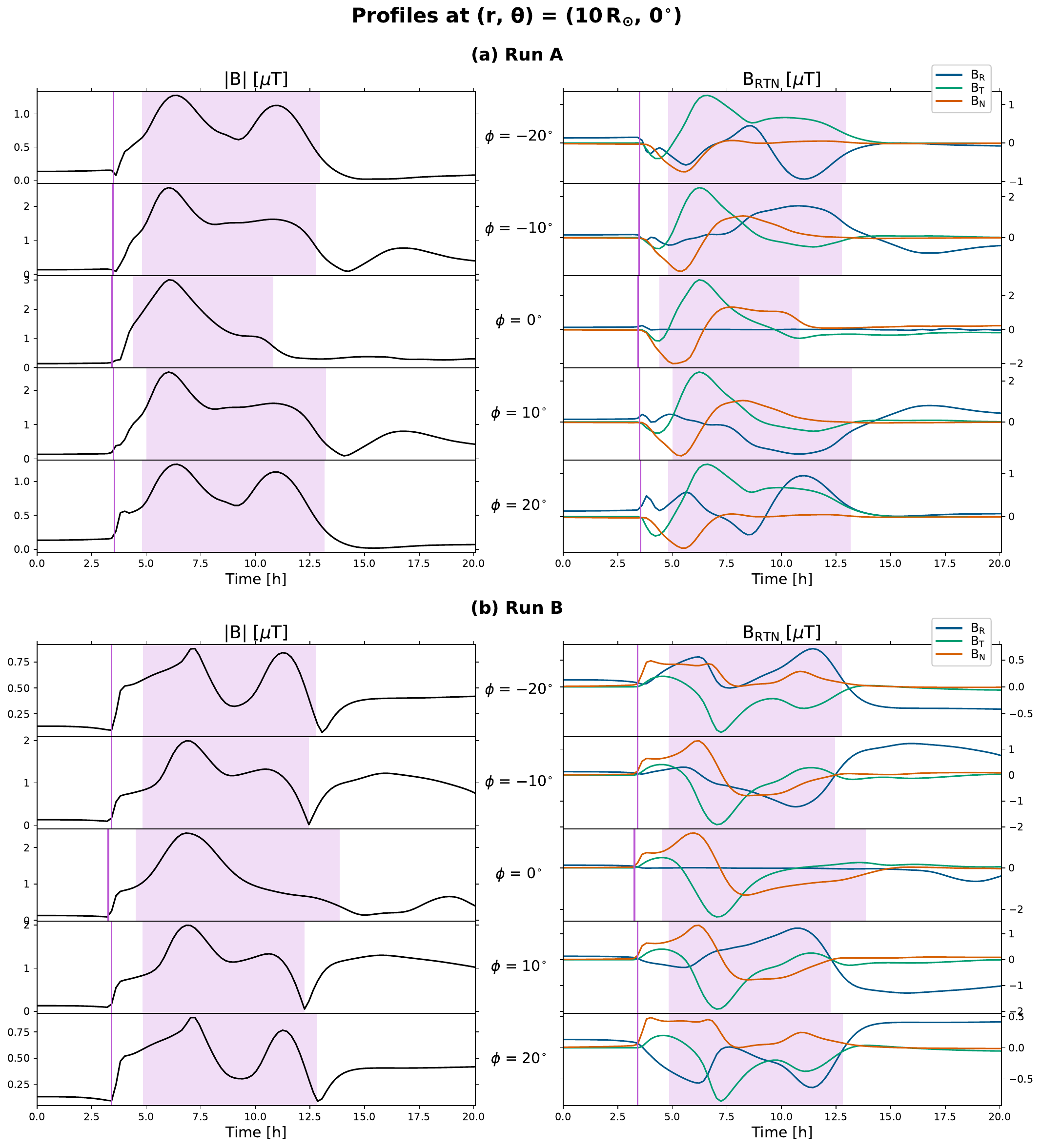}
\caption{Synthetic spacecraft profiles for fixed $(r, \theta) = (10\,R_{\odot}, 0^{\circ})$ at five longitudes ($-20^{\circ}$, $-10^{\circ}$, $0^{\circ}$, $10^{\circ}$, $20^{\circ}$) for (a) Run~A and (b) Run~B. The magnetic field magnitudes for each profile are shown to the left, whilst the corresponding magnetic field components in RTN coordinates are shown to the right. CME-driven sheaths are marked by vertical lines, whilst CME flux rope intervals are indicated by shaded areas. \label{fig:lonvar} }
\end{figure*}

The next series of synthetic spacecraft profiles that we analyse consists of probes at a fixed latitude in the solar equatorial plane, centred on evaluating longitudinal variations along the flux rope symmetry axis. As a representative radial distance that adequately samples the overall CME configuration in the corona after its early evolution stages \citep[e.g.,][]{kay2015b} and during its constant-propagation phase \citep[e.g.,][]{zhang2004}, we choose to inspect in-situ time series at $10\,R_{\odot}$. These profiles are shown in Figure~\ref{fig:lonvar}. We remark that, even if only magnetic field time series are shown, plasma data were also used in the determination of the sheath passage and ejecta boundaries (marked in violet as in Figure~\ref{fig:cmenose}). Additionally, we highlight that, although comprehensive 3D data are available from the MHD simulations, these profiles are analysed at this time based on synthetic in-situ measurements only, to treat interpretation of the structures under study in the same way as would be done for real spacecraft observations.

First of all, we note that the profiles become more complex and difficult to interpret with increasing sampling distance from the CME nose in both runs. At $\phi = {\pm}10^{\circ}$ from the apex, the magnitude of the $B_{R}$ component is comparable to the remaining two, but there are still a clear rotation in $B_{N}$ and a dominant pointing direction in $B_{T}$ that can lead to a correct visual determination of the intrinsic flux rope type---SWN for Run~A, NES for Run~B (see also Section~\ref{subsec:nose}). At $\phi = {\pm}20^{\circ}$ from the apex, on the other hand, there are several features that make assessments of the corresponding magnetic structure increasingly difficult. The field magnitudes are characterised by a double-peak trend, the $B_{R}$ component appears to display multiple rotations, and the $B_{N}$ component only rotates from the south (Run~A) or the north (Run~B) to approximately zero. \edit1{We note that such an asymmetric $B_{N}$ profile may be interpreted as the result of flux rope erosion \citep[e.g.,][]{dasso2006, lavraud2014}. Alternatively}, these ``double-rope'' signatures may be indicative of the passage through a flank of the main CME body and then a leg, as was demonstrated by \citet{owens2012} \edit1{and \citet{mostl2020}} via analytical modelling. However, it would not be possible to unambiguously determine a flux rope type or even a sense of chirality based on these measurements only, at least in terms of visual inspection of the data.

Additionally, we note in some profiles a considerable increase in the magnetic field magnitude, reflecting a corresponding increase in $B_{R}$, after the passage of what we defined as the CME ejecta, especially in the $\phi = {\pm}10^{\circ}$ profiles in both simulations and in the $\phi = {\pm}20^{\circ}$ profiles in Run~B. Given that both eruptions were simulated to propagate through a nearly-uniform background, it follows that any deviation from the ambient solar wind must be related to CME evolution---in this case, it is due to flows in the wake of the ejecta or to legs that remain connected to the Sun throughout the CME passage across the solar corona. Overall, apart from the larger $B_{R}$ increase after the ejecta measured in Run~B  (and, as expected, the opposite orientation of the two ropes), we do not find significant differences in the magnetic field behaviour encountered in the two runs.

\subsection{Latitudinal Variations at 10\,$R_{\odot}$} \label{subsec:latvar}

\begin{figure*}[p!]
\centering
\includegraphics[width=0.99\linewidth]{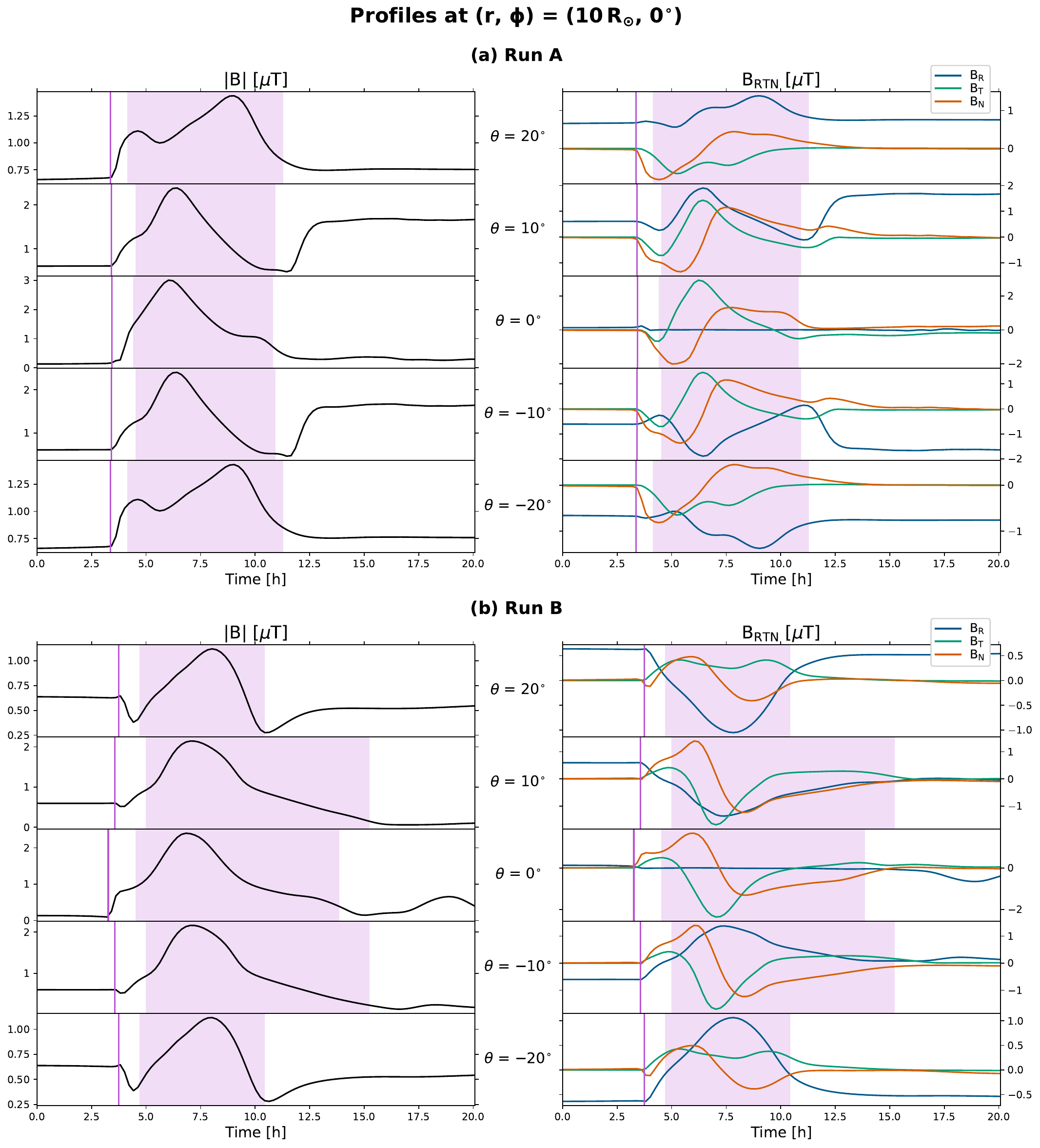}
\caption{Synthetic spacecraft profiles for fixed $(r, \phi) = (10\,R_{\odot}, 0^{\circ})$ at five latitudes ($20^{\circ}$, $10^{\circ}$, $0^{\circ}$, $-10^{\circ}$, $-20^{\circ}$) for (a) Run~A and (b) Run~B. The magnetic field magnitudes for each profile are shown to the left, whilst the corresponding magnetic field components in RTN coordinates are shown to the right. CME-driven sheaths are marked by vertical lines, whilst CME flux rope intervals are indicated by shaded areas. \label{fig:latvar} }
\end{figure*}

The next series of synthetic spacecraft profiles that we analyse consists of probes fixed at the longitude of the fictitious Sun--Earth line, centred on evaluating latitudinal variations in the direction perpendicular the flux rope symmetry axis. These profiles are shown in Figure~\ref{fig:latvar}, displayed again at a representative distance of $r = 10\,R_{\odot}$ and in the same format as Figure~\ref{fig:lonvar}. As in the previous case, we use a combination of magnetic field and plasma data to determine the times of sheath and ejecta passages.

Again, we note that the various time series become increasingly more complex with sampling distance from the CME nose in both runs. At $\theta = {\pm}10^{\circ}$ from the apex, the magnitude of the $B_{R}$ component is comparable to the remaining two; nevertheless, as in the $\phi = {\pm}10^{\circ}$ case, the distinct rotation in the $B_{N}$ component and the clear pointing direction in $B_{T}$ make visual identification of the corresponding flux rope type---SWN for Run~A, NES for Run~B---still rather straightforward. However, the $\theta = {\pm}20^{\circ}$ profiles display an interesting feature: Despite the $B_{R}$ component dominating in magnitude by a factor of ${\sim}2$ over the others, it is possible to clearly discern a rotation in $B_{N}$ and a single pointing direction in $B_{T}$. However, the flux rope type that would be inferred from visual inspection of these profiles is characterised by an opposite sense of chirality than that at the CME nose: SEN for Run~A and NWS for Run~B, which are both left-handed structures (note that both ropes were inserted with a positive helicity by design, see Section~\ref{sec:setup}). We remark that, from inspection of the $\theta = {\pm}20^{\circ}$ profiles alone, it would be possible to infer more or less unambiguously that the spacecraft is sampling the CME flank. However, there would be little reason to suspect that the chirality inferred from visual inspection is incorrect and that the underlying flux rope in fact possesses the opposite handedness.

In terms of differences between the two runs, we identify three main aspects. Strong radial flows in the wake of the CME are present only in Run~A at $\theta = {\pm}10^{\circ}$. In addition, the ejecta in Run~B tend to display a longer duration for $|\theta| \leq 10^{\circ}$ than their Run~A counterparts. Furthermore, the Run~B ejecta at $\theta = {\pm}20^{\circ}$ is bounded by low-field regions because of its quadrupolar eruption site (see Section~\ref{sec:setup}), resulting in a CME with opposite $B_{R}$ to the local coronal magnetic field. This leads to a CME characterised by a current boundary layer with a low-field region at the interface.

\subsection{Overall Magnetic Field Configuration at 15\,$R_{\odot}$} \label{subsec:lonlatvar}

\begin{figure*}[p!]
\centering
\includegraphics[width=0.99\linewidth]{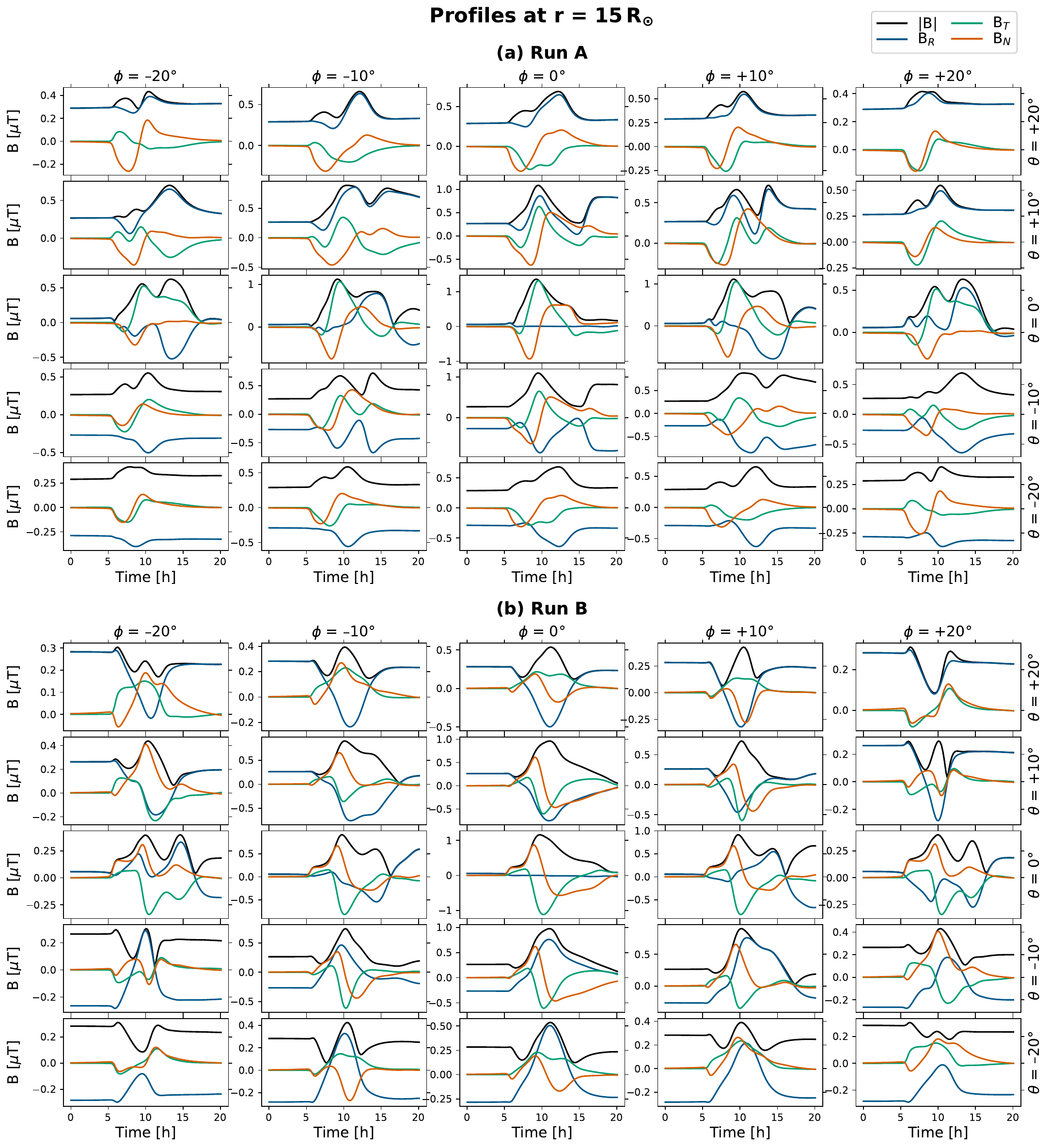}
\caption{Synthetic spacecraft profiles at $r = 15\,R_{\odot}$ for (a) Run~A and (b) Run~B. Each panel shows magnetic field magnitude as well as components in RTN coordinates for a different combination of ${\theta}$ and $\phi$. \label{fig:latlonvar} }
\end{figure*}

Finally, to better visualise the distribution of magnetic properties over the full CME angular extent at a given radial distance, we inspect the collective of in-situ profiles at a fixed radius. Figure~\ref{fig:latlonvar} shows such synthetic measurements at a representative distance of $r = 15\,R_{\odot}$, which corresponds to the approximate middle of our simulation domain (1--30\,$R_{\odot}$, see Section~\ref{sec:setup}). Within each panel ((a) for Run~A, (b) for Run~B), the central plot displays the encounter at the CME nose, and the surrounding time series feature profiles at successive ${\pm}10^{\circ}$ increments in $\theta$ and/or $\phi$.

As noted in the previous sections, it is clear that the $B_{R}$ component becomes increasingly more dominant with distance from the CME nose, which is to be expected based on theoretical flux rope models---where the magnitude of the radial field is related to the so-called impact parameter, i.e.\ the crossing distance from the central axis \citep[e.g.,][]{lepping1990, janvier2015}. In fact, we note that the contribution of $B_{R}$ to the total magnetic field grows faster in latitude than in longitude, which is consistent with the flux rope symmetry axis being roughly east--west oriented in the simulations investigated here \edit1{(in both the pre- and post-eruptive configurations)}. The profiles in either run---in particular $|B|$, $B_{T}$, and $B_{N}$---appear symmetric with respect to the ${\theta} = {\phi}$ line, with the exception of the $B_{R}$ time series changing sign, as expected of the idealised setup considered in this work.

The magnetic field magnitude profiles at the CME nose conform to the ``ideal'' picture of an expanding flux rope, characterised by a decrease in field strength from the front to the back of the ejecta \citep[e.g.,][]{nieveschinchilla2018a}, and become more complex with distance from the apex. The distribution of trends in $|B|$ does not evolve in the same manner between Run~A (Figure~\ref{fig:latlonvar}(a)) and Run~B (Figure~\ref{fig:latlonvar}(b))---for example, the profiles at ($\theta$, $\phi$) = (${\pm}10^{\circ}$, ${\pm}10^{\circ}$)  show a nose-like behaviour in Run~B, but feature multiple peaks and an increasing trend in Run~A. Of particular note are the profiles at ($\theta$, $\phi$) = (${\pm}20^{\circ}$, ${\pm}20^{\circ}$): whilst in Run~A it is possible to identify the CME passage as a (single- or double-peaked) increase in $|B|$, in Run~B the ejecta appears as a ``depression'' over the ambient coronal magnetic fields. This is in contrast with the ``classic'' picture of CME ejecta manifesting as a field enhancement over the background solar wind, but we note that Parker Solar Probe measurements at short heliocentric distances have shown several cases of a CME flank and/or leg encounter with field magnitudes comparable to the ambient medium \citep[e.g.,][]{getachew2022, mccomas2023, braga2024}.

We also consider the flux rope chirality and type that would be inferred via visual inspection of the collective of profiles. As mentioned in Section~\ref{subsec:nose}, in-situ measurements at the CME nose allow us to retrieve two right-handed flux ropes with oppositely-oriented axes, of type SWN for Run~A and NES for Run~B. We noted by examining profiles away from the nose in latitude (Section~\ref{subsec:latvar}) and longitude (Section~\ref{subsec:lonvar}) that, at selected locations, estimations of a well-defined flux rope type either become ambiguous or lead to determination of a left-handed rope, of opposite chirality to the pre-eruptive structure that was designed (see Section~\ref{sec:setup}). Here, we further evaluate the variability of the recovered magnetic field configurations across the whole CME extent. Amongst the profiles at $\theta = {\pm}20^{\circ}$, we note that in either run all but one location per given latitude display $B_{T}$ fields that are predominantly of opposite sign than at the nose. The $B_{N}$ component, on the other hand, features different characteristics depending on the CME under analysis: In Run~A, all profiles still show a rotation from south to north (as at the apex), whilst in Run~B a clear transition from north to south is only present at $\phi = 0^{\circ}$, i.e.\ directly above/below the nose---even more so, the time series at ($\theta$, $\phi$) = ($20^{\circ}$, $20^{\circ}$)  and ($-20^{\circ}$, $-20^{\circ}$) appear to instead turn from south to north. 

The profiles at $\theta = {\pm}10^{\circ}$ show a variety of trends in the $B_{T}$ and $B_{N}$ series: Either component rotates (from negative to positive or vice versa) in only approximately half of the time series, with the remaining sets of measurements displaying a clear ($B_{T}$ or $B_{N}$) direction that decreases to near-zero in the trailing portion of the ejecta. We note that different longitudes feature different combinations of (rotating and/or non-rotating) $B_{T}$ and $B_{N}$ components, yielding inconsistent estimates of the magnetic configuration of the underlying flux rope(s). Hence, at least according to visual inspection of various time series displayed in Figure~\ref{fig:latlonvar}, it is clear that crossings at multiple locations through a given CME would result in retrieval of a set of flux ropes that may suggest not only different orientations and physical properties, but also opposite chiralities. 


\section{Analysis and Interpretation} \label{sec:interpretation}

In the previous section, we have provided an overview of the different profiles at the synthetic observers for both runs and highlighted the main characteristics and patterns that we have encountered. Here, we focus on interesting features that emerged from visual inspection of the collective in-situ data and investigate them further in the context of two main aspects: variability of the inferred CME morphology and magnetic structure based on spacecraft crossing location (Section~\ref{subsec:chirality}) and variability of intrinsic CME structure and evolution based on source region topology (Section~\ref{subsec:topology}).

\subsection{Determination of CME Structure} \label{subsec:chirality}

Throughout Section~\ref{sec:results}, we have determined the flux rope type that would be inferred at each spacecraft crossing based uniquely on visual inspection of the magnetic field data. Here, we aim to quantify how the local magnetic configuration varies across the CME with the aid of flux rope fitting techniques. As noted in Section~\ref{subsec:nose}, we did not find significant variations with radial distance along the CME nose, except for the flux rope showing signatures of expansion. After verifying that the same holds true for crossings at other locations \edit1{(i.e., that the CMEs expands largely self-similarly)}, we set to investigate how the inferred flux rope type changes over the CME cross-section at a given heliocentric altitude---here, we focus on $10\,R_{\odot}$ as an exemplary distance, which also approximates Parker Solar Probe's closest perihelion of $9.86\,R_{\odot}$ \citep[e.g.,][]{raouafi2023}. Given the symmetries identified with respect to the ${\theta} = {\phi}$ line (see Section~\ref{subsec:lonlatvar}), we select a few sample locations that are representative of the variety of profiles encountered, examining thus five profiles per run. The flux rope model that we employ here is the expansion-modified force-free cylinder \citep[e.g.,][]{farrugia1993, yu2022} in the same implementation as that of \citet{palmerio2025}, which considers the solar wind speed as one of the parameters constraining the fit alongside the magnetic field magnitude and components. With respect to the classic constant-$\alpha$ force-free solution \citep[e.g.,][]{burlaga1988, lepping1990}, this model includes an additional `expansion time' parameter that describes the self-similar radial expansion of the flux rope.

\begin{figure*}[th!]
\centering
\includegraphics[width=0.99\linewidth]{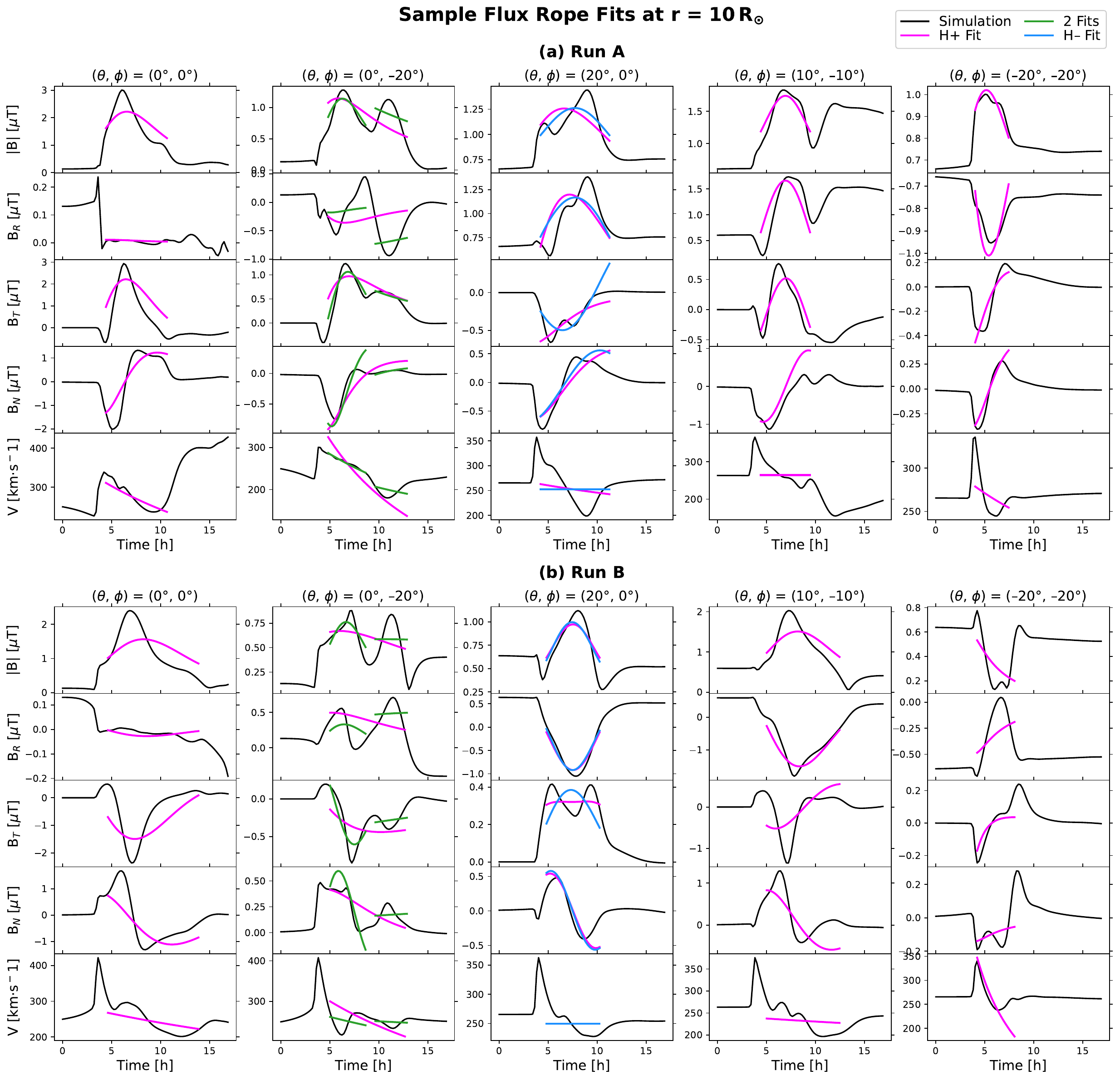}
\caption{Fits with the expansion-modified force-free cylinder flux rope model applied to representative in-situ profiles for (a) Run~A and (b) Run~B. Simulated data are shown in black, single-rope fits considering a positive helicity sign are shown in magenta, two-rope fits performed for the ($\theta$, $\phi$) = ($0^{\circ}$, $-20^{\circ}$) profiles are shown green, and single-rope fits assuming a negative helicity sign performed for the ($\theta$, $\phi$) = ($-20^{\circ}$, $-20^{\circ}$) profiles are shown in blue. \label{fig:frfits} }
\end{figure*}

Results from the flux rope fitting procedure are summarised in Figure~\ref{fig:frfits} and Table~\ref{tab:frfits}. For each profile, we perform one fit spanning the entire identified flux rope interval assuming a right-handed chirality (magenta curves)---these are based on both visual inspection and the known, pre-eruptive magnetic configuration described in Section~\ref{sec:setup}. For the profiles at ($\theta$, $\phi$) = ($0^{\circ}$, $-20^{\circ}$), considered here as representative cases of double-peak profiles (first discussed in Section~\ref{subsec:lonvar}), we additionally perform two separate fits (green curves), each corresponding to one of the increases in magnetic field magnitude. Finally, for the profiles at ($\theta$, $\phi$) = ($20^{\circ}$, $0^{\circ}$), we additionally perform one fit that assumes a left-handed chirality (blue curves), as may have been assumed from visual inspection of the magnetic field components (see Section~\ref{subsec:latvar}). The results presented in Figure~\ref{fig:frfits} and Table~\ref{tab:frfits} show that not only the magnetic configuration of the retrieved flux rope, but also the performance of the algorithm itself are highly dependent on the spacecraft crossing location.

\begin{table*}[t!]
\caption{Flux rope fitting results at selected virtual spacecraft crossings. \label{tab:frfits}}
\centering
\renewcommand{\arraystretch}{1}
\begin{tabularx}{\linewidth}{lcccccccccc}
\toprule
 & \textbf{H} & $\mathbf{({\Theta}_{0}, {\Phi}_{0})}$ & $\mathbf{p_{0}}$ & $\mathbf{B_{0}}$ & $\mathbf{\tau_{0}}$ & $\mathbf{R_{0}}$ & $\mathbf{R_{1}}$ & $\mathbf{V_\mathrm{\mathbf{exp}}}$ & $\mathbf{\chi^{2}_\mathrm{\mathbf{dir}}}$ & $\mathbf{\chi^{2}_\mathrm{\mathbf{mag}}}$\\
\midrule
\textsc{\textbf{Run~A}}\\  
\cmidrule(r{10pt}){1-1}
(${\theta}$, ${\phi}$) = ($0^{\circ}$, $0^{\circ}$) & +1 & ($15^{\circ}$, $90^{\circ}$) & 0.01 & 3.8~$\mu$T & 23~h & $4.4\,R_{\odot}$ & $6.5\,R_{\odot}$ & 37~km$\cdot$s$^{-1}$ & 0.115 & 0.044\\
(${\theta}$, ${\phi}$) = ($0^{\circ}$, $-20^{\circ}$) & +1 & ($-8^{\circ}$, $107^{\circ}$) & $-0.09$ & 3.3~$\mu$T & 9~h & $4.6\,R_{\odot}$ & $10.9\,R_{\odot}$ & 94~km$\cdot$s$^{-1}$ & 0.282 & 0.046\\
(${\theta}$, ${\phi}$) = ($0^{\circ}$, $-20^{\circ}$)$^\mathrm{a}$ & +1 & ($-20^{\circ}$, $83^{\circ}$) & $-0.34$ & 2.0~$\mu$T & 21~h & $2.8\,R_{\odot}$ & $3.9\,R_{\odot}$ & 25~km$\cdot$s$^{-1}$ & 0.140 & 0.008\\
(${\theta}$, ${\phi}$) = ($0^{\circ}$, $-20^{\circ}$)$^\mathrm{b}$ & +1 & ($1^{\circ}$, $161^{\circ}$) & $0.99$ & 3.3~$\mu$T & 13~h & $5.2\,R_{\odot}$ & $10.3\,R_{\odot}$ & 78~km$\cdot$s$^{-1}$ & 0.297 & 0.041\\
(${\theta}$, ${\phi}$) = ($20^{\circ}$, $0^{\circ}$) & +1 & ($-4^{\circ}$, $23^{\circ}$) & $0.77$ & 2.2~$\mu$T & 34~h & $2.8\,R_{\odot}$ & $3.7\,R_{\odot}$ & 16~km$\cdot$s$^{-1}$ & 0.122 & 0.016\\
(${\theta}$, ${\phi}$) = ($20^{\circ}$, $0^{\circ}$) & $-1$ & ($30^{\circ}$, $277^{\circ}$) & $-0.77$ & 1.9~$\mu$T & 1$\times$10$^{5}$~h & $7.1\,R_{\odot}$ & $7.1\,R_{\odot}$ & 0~km$\cdot$s$^{-1}$ & 0.078 & 0.009\\
(${\theta}$, ${\phi}$) = ($10^{\circ}$, $-10^{\circ}$) & +1 & ($-2^{\circ}$, $64^{\circ}$) & $0.62$ & 2.3~$\mu$T & 3$\times$10$^{8}$~h & $3.9\,R_{\odot}$ & $3.9\,R_{\odot}$ & 0~km$\cdot$s$^{-1}$ & 0.117 & 0.005\\
(${\theta}$, ${\phi}$) = ($-20^{\circ}$, $-20^{\circ}$) & +1 & ($-12^{\circ}$, $152^{\circ}$) & $-0.81$ & 2.0~$\mu$T & 19~h & $2.1\,R_{\odot}$ & $2.8\,R_{\odot}$ & 21~km$\cdot$s$^{-1}$ & 0.235 & 0.001\\
\midrule
\textsc{\textbf{Run~B}}\\  
\cmidrule(r{10pt}){1-1}
(${\theta}$, ${\phi}$) = ($0^{\circ}$, $0^{\circ}$) & +1 & ($-32^{\circ}$, $269^{\circ}$) & 0.00 & 2.1~$\mu$T & 50~h & $5.9\,R_{\odot}$ & $7.5\,R_{\odot}$ & 23~km$\cdot$s$^{-1}$ & 0.168 & 0.107\\
(${\theta}$, ${\phi}$) = ($0^{\circ}$, $-20^{\circ}$) & +1 & ($30^{\circ}$, $277^{\circ}$) & $0.89$ & 1.4~$\mu$T & 22~h & $10.8\,R_{\odot}$ & $16.9\,R_{\odot}$ & 94~km$\cdot$s$^{-1}$ & 0.192 & 0.073\\
(${\theta}$, ${\phi}$) = ($0^{\circ}$, $-20^{\circ}$)$^\mathrm{a}$ & +1 & ($41^{\circ}$, $267^{\circ}$) & $0.52$ & 1.1~$\mu$T & 44~h & $2.8\,R_{\odot}$ & $3.3\,R_{\odot}$ & 12~km$\cdot$s$^{-1}$ & 0.109 & 0.011\\
(${\theta}$, ${\phi}$) = ($0^{\circ}$, $-20^{\circ}$)$^\mathrm{b}$ & +1 & ($-18^{\circ}$, $31^{\circ}$) & $1.00$ & 1.1~$\mu$T & 122~h & $29.9\,R_{\odot}$ & $33.0\,R_{\odot}$ & 47~km$\cdot$s$^{-1}$ & 0.096 & 0.125\\
(${\theta}$, ${\phi}$) = ($20^{\circ}$, $0^{\circ}$) & +1 & ($0^{\circ}$, $200^{\circ}$) & $-0.53$ & 1.2~$\mu$T & 1$\times$10$^{9}$~h & $1.4\,R_{\odot}$ & $1.4\,R_{\odot}$ & 0~km$\cdot$s$^{-1}$ & 0.037 & 0.018\\
(${\theta}$, ${\phi}$) = ($20^{\circ}$, $0^{\circ}$) & $-1$ & ($0^{\circ}$, $180^{\circ}$) & $-0.35$ & 1.2~$\mu$T & 187~h & n/a & n/a & 0~km$\cdot$s$^{-1}$ & 0.045 & 0.014\\
(${\theta}$, ${\phi}$) = ($10^{\circ}$, $-10^{\circ}$) & +1 & ($-13^{\circ}$, $196^{\circ}$) & $-0.30$ & 2.0~$\mu$T & 60~h & $1.6\,R_{\odot}$ & $2.0\,R_{\odot}$ & 5~km$\cdot$s$^{-1}$ & 0.112 & 0.046\\
(${\theta}$, ${\phi}$) = ($-20^{\circ}$, $-20^{\circ}$) & +1 & ($-15^{\circ}$, $181^{\circ}$) & $0.33$ & 7.5~$\mu$T & 2~h & $0.7\,R_{\odot}$ & $4.3\,R_{\odot}$ & 87~km$\cdot$s$^{-1}$ & 0.212 & 0.090\\
\bottomrule
\end{tabularx}
\vspace*{.1in}
\begin{tablenotes}
\item \emph{Notes.} The reported parameters are, from left to right: helicity sign (H), flux rope axis direction (${\Theta}_{0}, {\Phi}_{0}$), impact parameter normalised over the flux rope radius ($p_{0}$), magnetic field magnitude at the axis ($B_{0}$), expansion time ($\tau_{0}$), initial radius ($R_{0}$), final radius ($R_{1}$), expansion speed ($V_\mathrm{exp}$), error measure on the magnetic field components ($\chi^{2}_\mathrm{dir}$), and error measure on the magnetic field magnitude normalised by its maximum fitted value ($\chi^{2}_\mathrm{mag}$). The `a' and `b' superscripts denote, respectively, the first and second portion of double-peaked profiles when fitted separately.
\end{tablenotes}
\end{table*}

First, we note that all fits appear consistent with a generally low-inclination (${\Theta}_{0}\lesssim 45^{\circ}$) flux rope in both simulations, in agreement with the pre-eruptive structures imposed at the Sun---specifically, SWN for Run~A and NES for Run~B (see Section~\ref{sec:setup}). However, the exact axis direction, with (${\Theta}_{0}, {\Phi}_{0}$) expected to lie close to ($0^{\circ}$, $90^{\circ}$) for Run~A and ($0^{\circ}$, $270^{\circ}$) for Run~B, is retrieved adequately only for the profiles at (${\theta}$, ${\phi}$) = ($0^{\circ}$, $0^{\circ}$) and ($0^{\circ}$, $-20^{\circ}$) in both runs, as well as ($10^{\circ}$, $-10^{\circ}$) in Run~A \citep[considering uncertainties of the order of ${\sim}30^{\circ}$ in ${\Theta}_{0}$ and ${\Phi}_{0}$; e.g.,][]{lynch2005a, alhaddad2013}. This again demonstrates that the performance of flux rope fitting techniques deteriorates with increasing distance from the flux rope axis \citep[e.g.,][]{lepping2003, riley2004}, which in these simulations lies along the ${\theta}=0^{\circ}$ plane. Interestingly, the fitting results at the CME nose are not associated with the smallest errors across all cases. This likely reflects the fact that at low coronal heights CMEs can undergo rapid expansion \citep[e.g.,][]{veronig2018, zhuang2022}, leading to highly peaked and asymmetric profiles that deviate from the smooth, idealised structures usually assumed in analytical flux rope models. We also note that the Run~A fits are generally associated with lower errors and yield more physically viable results, particularly in terms of the flux rope radius, keeping in mind that the synthetic in-situ encounters are extracted at $r = 10\,R_{\odot}$. This may be attributed to the more ``classic'' profiles obtained in the case of the bipolar eruption (Run~A), in contrast to the quadrupolar configuration (Run~B), which produces more complex magnetic field signatures, including cases where the CME appears as a depression relative to the ambient environment (see also Section~\ref{subsec:lonlatvar} and Figure~\ref{fig:latlonvar}). Overall, while the fits are of variable quality, previous modelling \citep[e.g.,][]{lynch2022, lynch2025} as well as observational \citep[e.g.,][]{braga2024, davies2024} studies have shown that flux rope fitting techniques can be reasonably applied to in-situ measurements \edit1{of CMEs encountered in the corona}, provided that the appropriate caveats are taken into account.

The profiles displaying a double peak in the magnetic field magnitude---taking the (${\theta}$, ${\phi}$) = ($0^{\circ}$, $-20^{\circ}$) crossings as an example---yield results that are more mutually consistent, and more consistent with the nose profiles, when either the full interval (including both peaks) or only the first enhancement is fitted. In contrast, fits restricted to the second field increase alone deviate substantially from the overall flux rope configuration (and, in the case of Run~B, feature an unrealistic flux rope radius of ${\sim}30\,R_{\odot}$). This behaviour is plausible if the second enhancement is interpreted as a crossing closer to the CME leg, where the local flux rope axis is expected to become more aligned with the radial direction than in the front portion of the ejecta, consistent with the retrieved orientations. The full-interval fit obtained for the Run~B CME, however, results in an unrealistic size of the structure, with the fitted radius approaching $11\,R_{\odot}$ despite the CME being first encountered at $10\,R_{\odot}$. Overall, these results suggest that fitting only the first coherent magnetic field rotation yields configurations that are more representative of the global CME structure. 

\begin{figure*}[th!]
\centering
\includegraphics[width=0.495\linewidth]{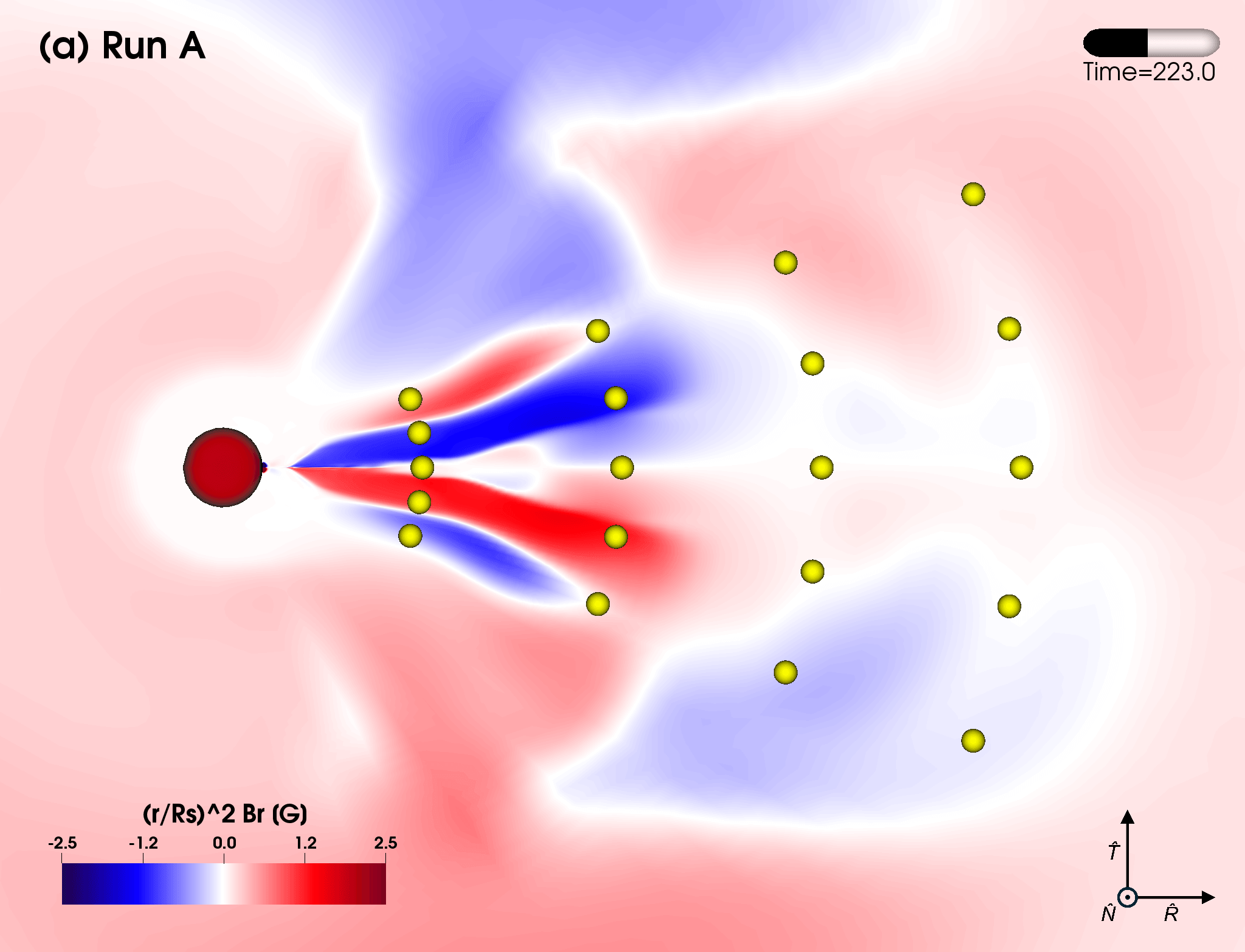}
\includegraphics[width=0.495\linewidth]{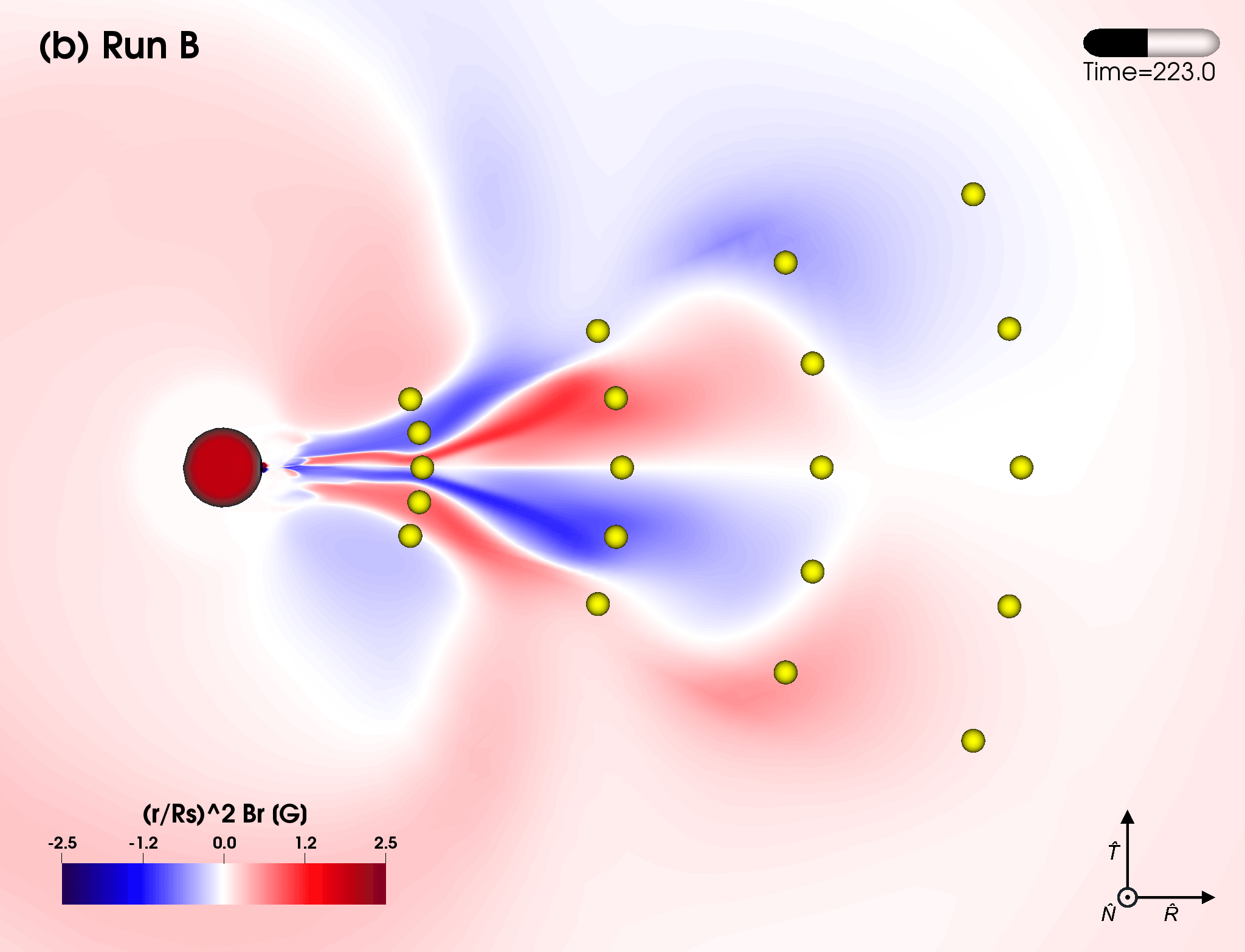}
\caption{Snapshot (at $t \simeq 9.5$~h) of the scaled radial magnetic field component ($B_\mathrm{R}$) for (a) Run~A and (b) Run~B projected onto the equatorial plane as seen from above the solar north pole. The synthetic spacecraft that cross the equatorial plane are represented by yellow spheres. The animated version of this figure shows the evolution of $B_\mathrm{R}$ throughout the temporal domain of the simulations (0--20~h).\\
(An animation of this figure is available.)
\label{fig:brslice} }
\end{figure*}

To verify whether these double-peak profiles correspond to the in-situ signature of a ``double crossing'' of the flux rope core fields---from the frontal portion of the CME into one of its legs---we examine equatorial slices of the $B_\mathrm{R}$ component, shown in Figure~\ref{fig:brslice}. Both the figure and its animated version clearly demonstrate that longitudes away from the hypothetical Sun--Earth line (${\phi} = 0^{\circ}$) become immersed, for varying portions of the simulation, in strong radial magnetic field regions associated with the CME legs and related large-scale magnetic structures. Such crossing geometries were investigated in detail by \citet{owens2012} using an analytical CME model, where they were termed ``double flux rope signatures''. In this scenario, the spacecraft crosses the flux rope axis twice owing to the intrinsic curvature of the CME axis combined with the particular trajectory of the observer through the structure. \citet{owens2012} further showed that fitting the two portions of the profile separately can yield satisfactory estimates of the local flux rope orientation(s), while also cautioning that, in extreme cases, such signatures may be difficult to distinguish from CME--CME interaction events.

\begin{figure*}[th!]
\centering
\includegraphics[width=0.495\linewidth]{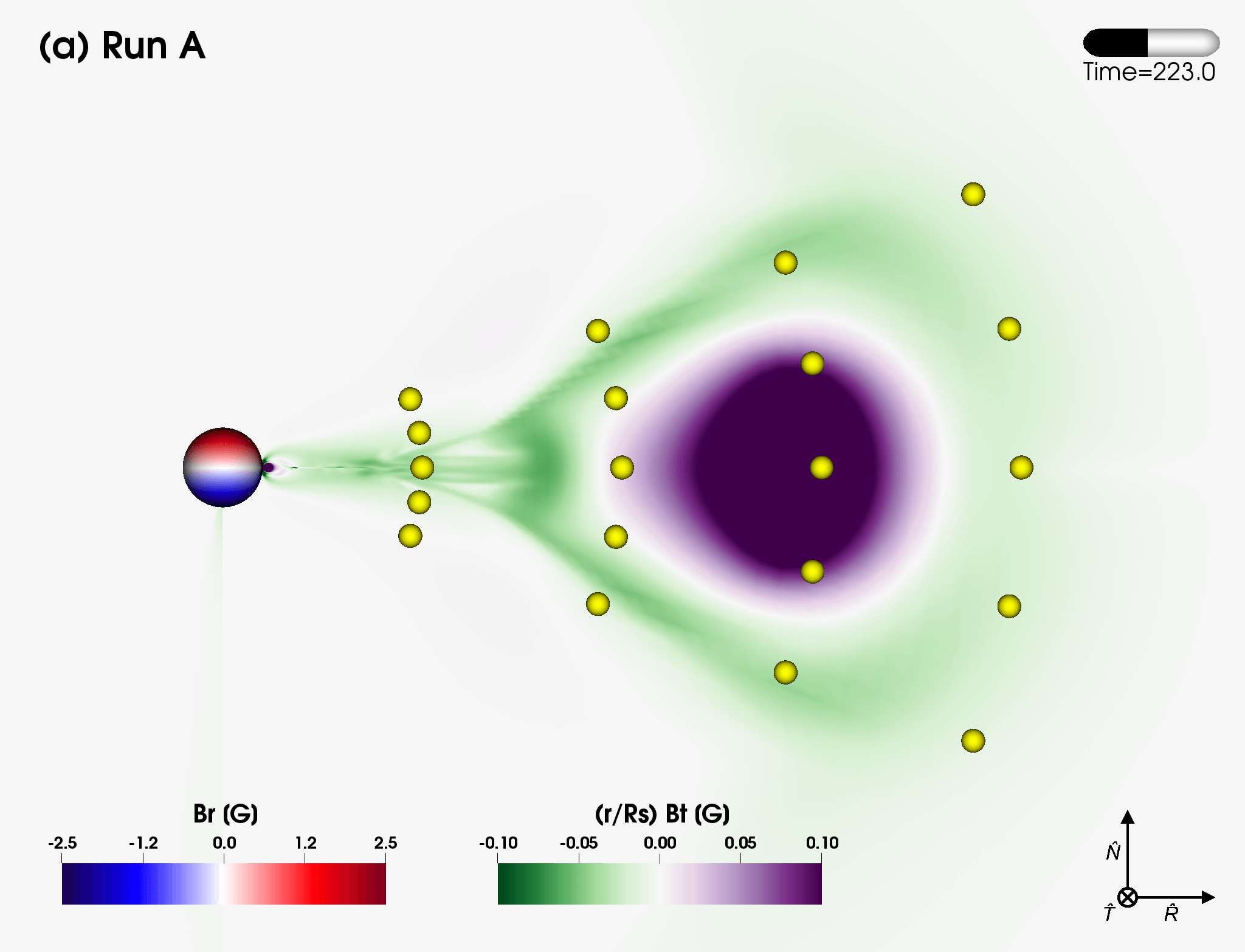}
\includegraphics[width=0.495\linewidth]{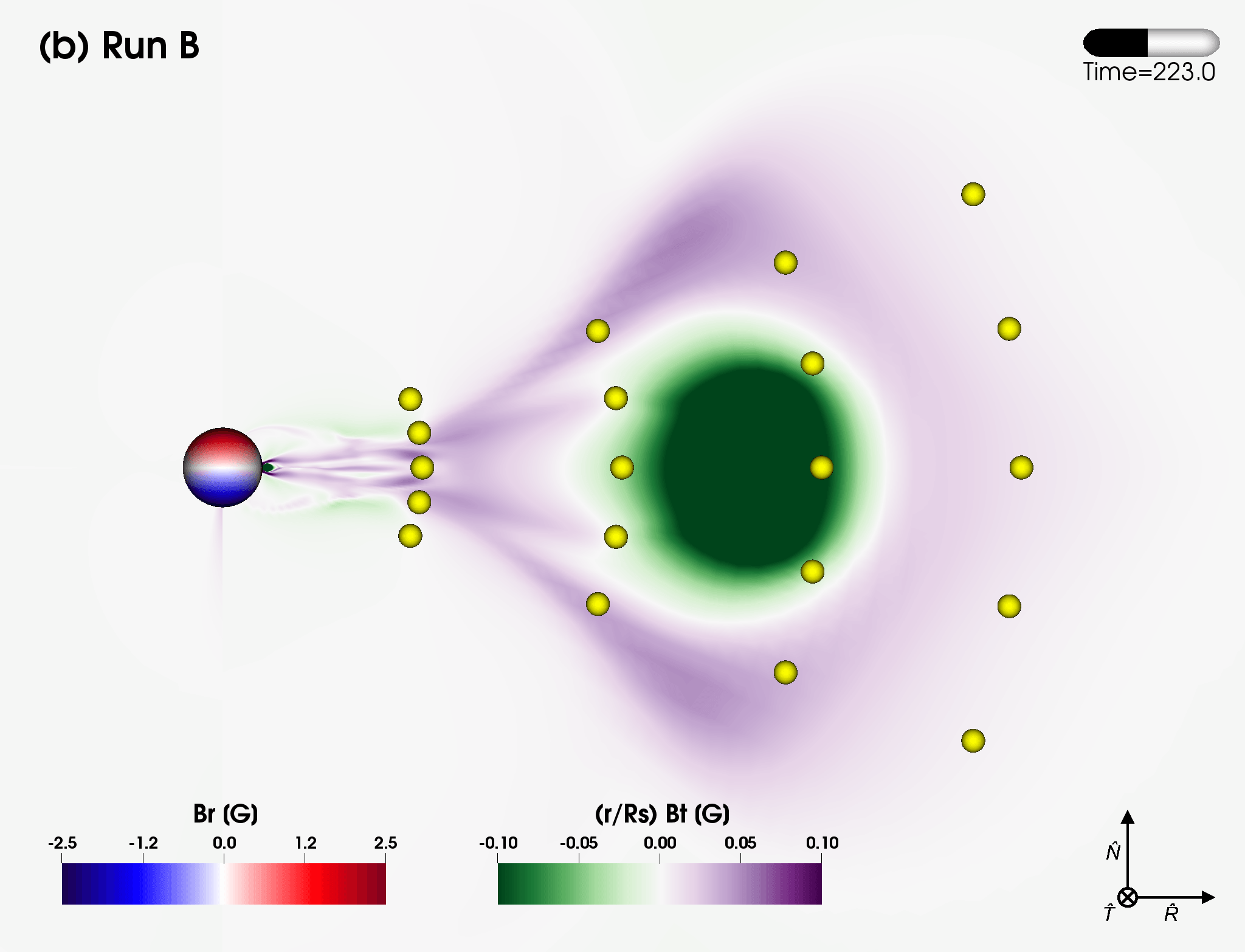}
\caption{Snapshot (at $t \simeq 9.5$~h) of the scaled tangential magnetic field component ($B_\mathrm{T}$) for (a) Run~A and (b) Run~B projected onto the meridional plane at $\phi = 0^{\circ}$. The synthetic spacecraft that cross the meridional plane are represented by yellow spheres. The animated version of this figure shows the evolution of $B_\mathrm{T}$ throughout the temporal domain of the simulations (0--20~h).\\
(An animation of this figure is available.)
\label{fig:btslice} }
\end{figure*}

\begin{figure*}[th!]
\centering
\includegraphics[width=0.495\linewidth]{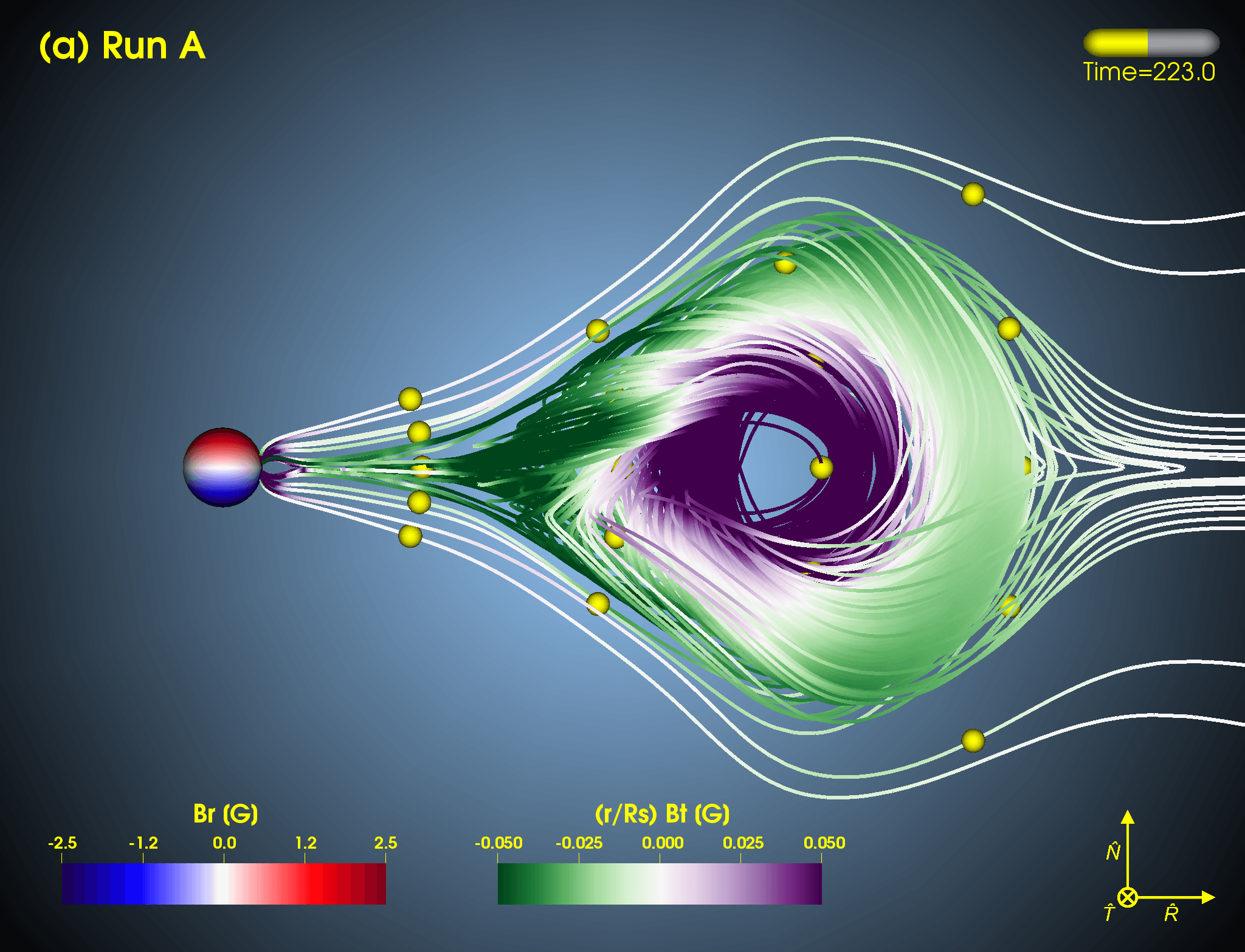}
\includegraphics[width=0.495\linewidth]{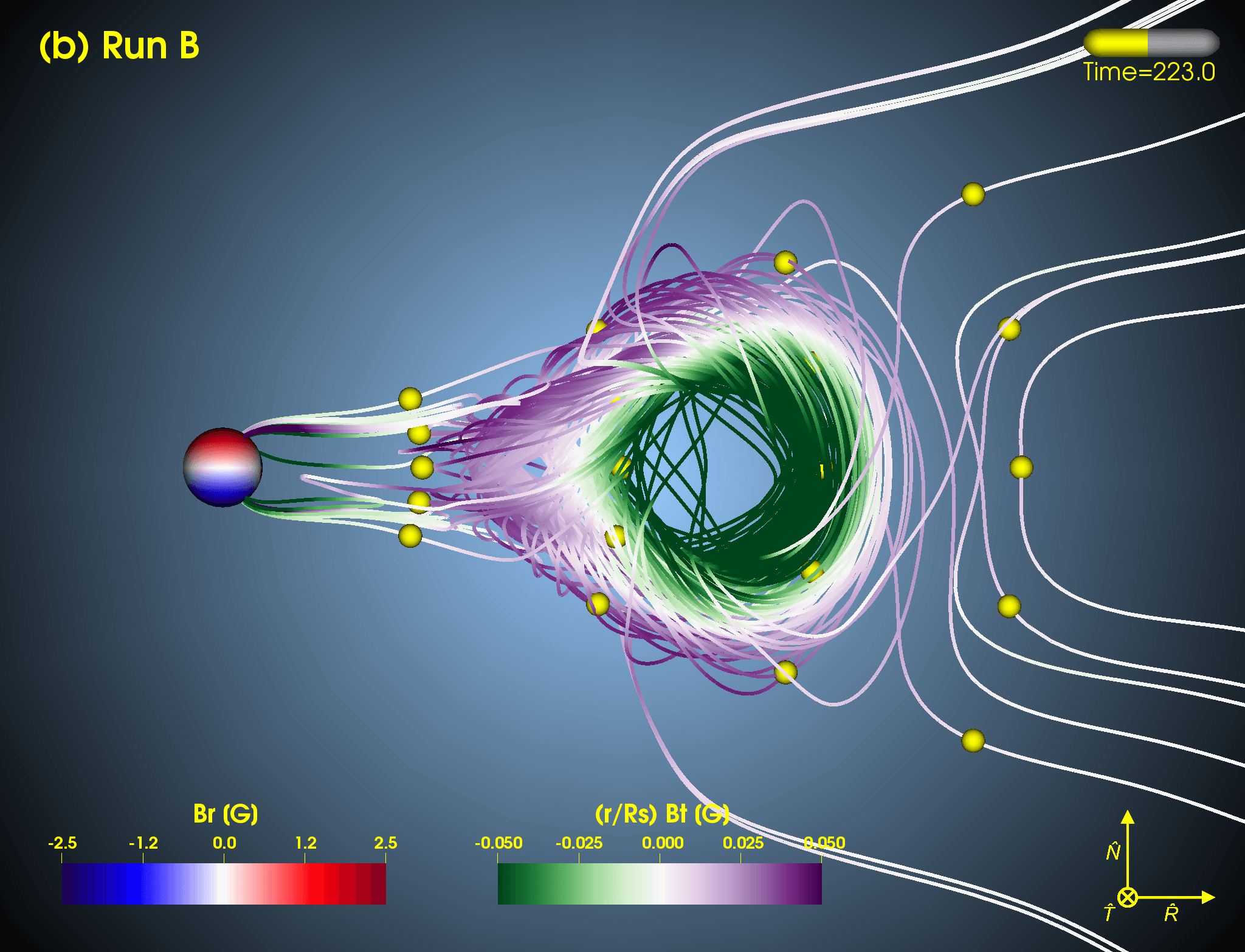}\\
\vspace*{.03in}
\includegraphics[width=0.495\linewidth]{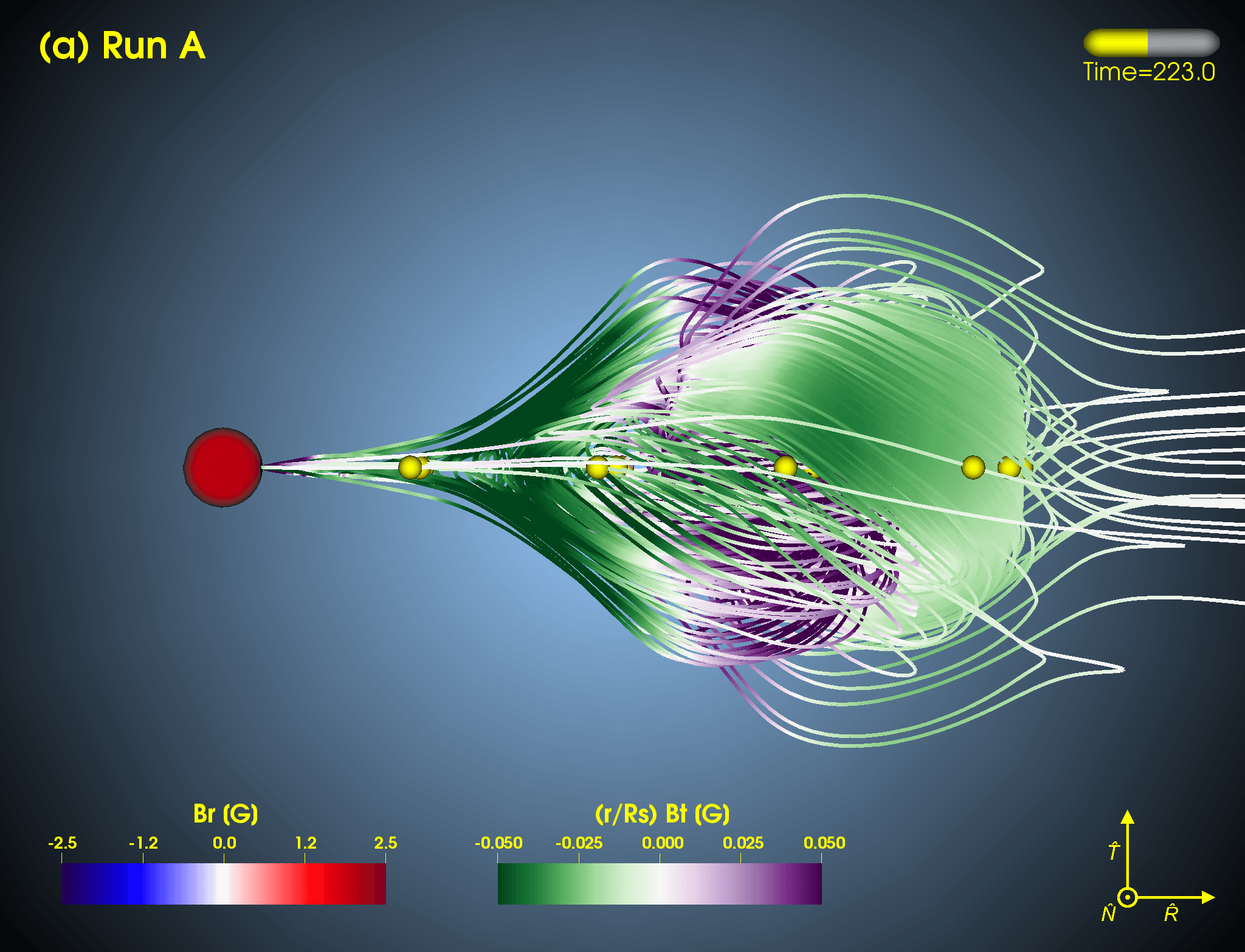}
\includegraphics[width=0.495\linewidth]{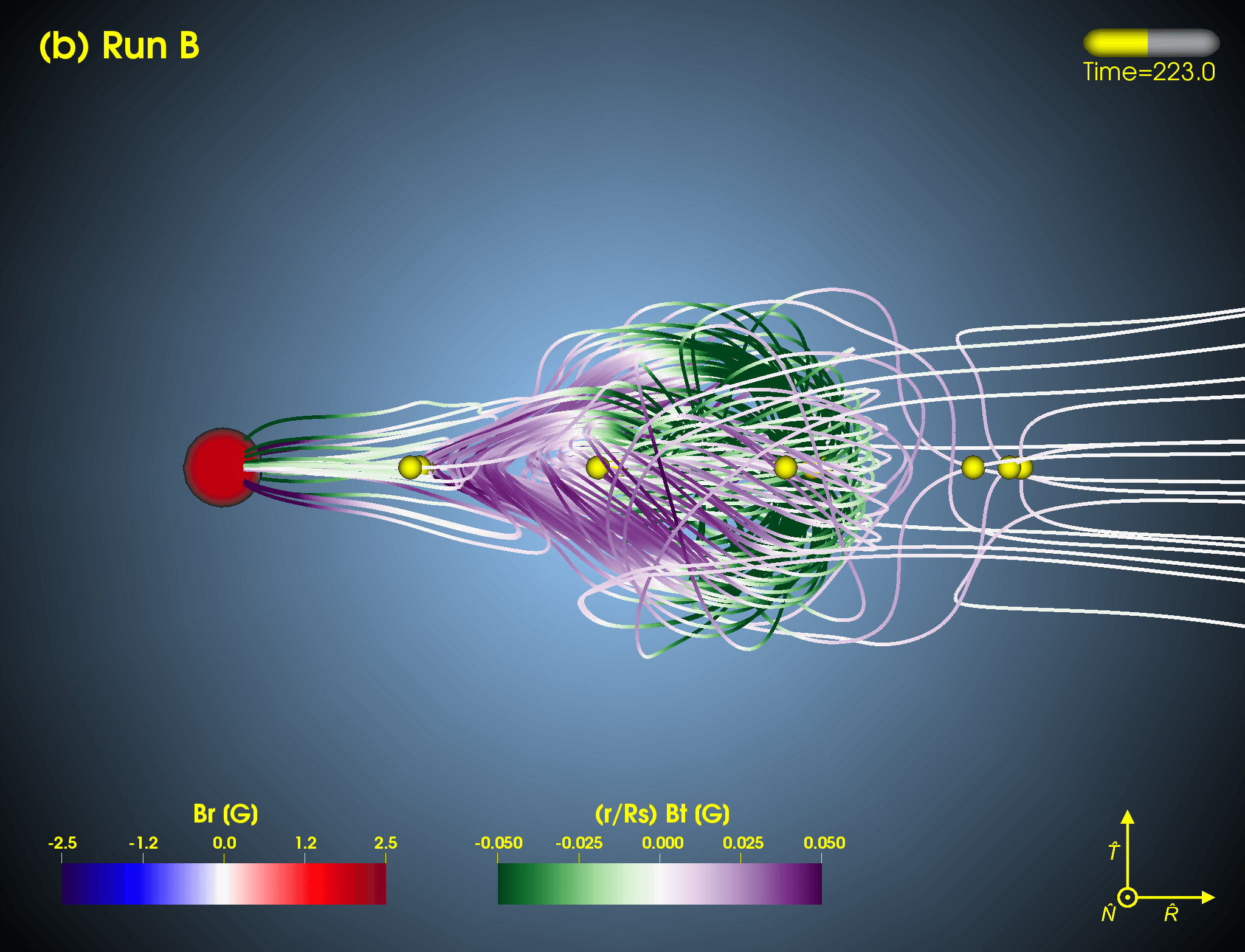}
\caption{Snapshot (at $t \simeq 9.5$~h) of the magnetic field lines connected to the 20 synthetic spacecraft located at $\phi=0^{\circ}$ for (a) Run~A and (b) Run~B projected onto (top) the meridional plane and (bottom) the equatorial plane. The synthetic spacecraft that cross the meridional plane are represented by yellow spheres. The field lines are coloured by the (scaled) magnitude and sign of the $B_{T}$ component. The animated version of this figure shows the evolution of these 20 field lines throughout the temporal domain of the simulations (0--20~h).\\
(An animation of this figure is available.)
\label{fig:fieldlines} }
\end{figure*}

Turning now to the fits at (${\theta}$, ${\phi}$) = ($20^{\circ}$, $0^{\circ}$), we find that they are largely inconclusive with respect to the helicity sign of the underlying flux rope, with comparably poor results obtained for both right- and left-handed configurations in the two simulations. In the case of Run~B, the retrieved flux rope axis is nearly radial, a geometry generally regarded as either indicative of an unreliable fit or corresponding to a leg crossing \citep[e.g.,][]{marubashi1997}. We therefore conclude that, although visual inspection alone would likely favour a left-handed interpretation for these profiles (see Section~\ref{subsec:latvar}), flux rope fitting yields a considerably more ambiguous result---likely arising from the dominant $B_\mathrm{R}$ component, which is often de-emphasised or neglected during qualitative inspection of in-situ data. To investigate the origin of the oppositely directed $B_{T}$ component present in most profiles at $\theta = {\pm}20^{\circ}$ (see also Figure~\ref{fig:latlonvar}), we examine meridional slices at ${\phi} = 0^{\circ}$, shown in Figure~\ref{fig:btslice}. The figure reveals that the strong central flux rope core---directed westwards in Run~A and eastwards in Run~B---is surrounded by a weaker magnetic field of opposite polarity. This configuration is already present during the early stages of the eruption, as can be seen in the animated version of the figure. Such a feature may arise from magnetic field draping around the expanding CME ejecta \citep[e.g.,][]{manchester2004a, liu2011} and/or from the surface-current and return-current systems surrounding erupting flux ropes in the corona, as discussed in detail for our runs by \citet[][see their Figure~7]{bennun2023}, and more generally by \citet{torok2014} as well as \citet{titov2022}. Reconnection-driven restructuring and erosion of the outer flux surfaces may further contribute to the formation of oppositely directed magnetic layers around the core flux rope \citep[e.g.,][]{hosteaux2018, wyper2024}. 

\begin{figure*}[th!]
\centering
\includegraphics[width=0.495\linewidth]{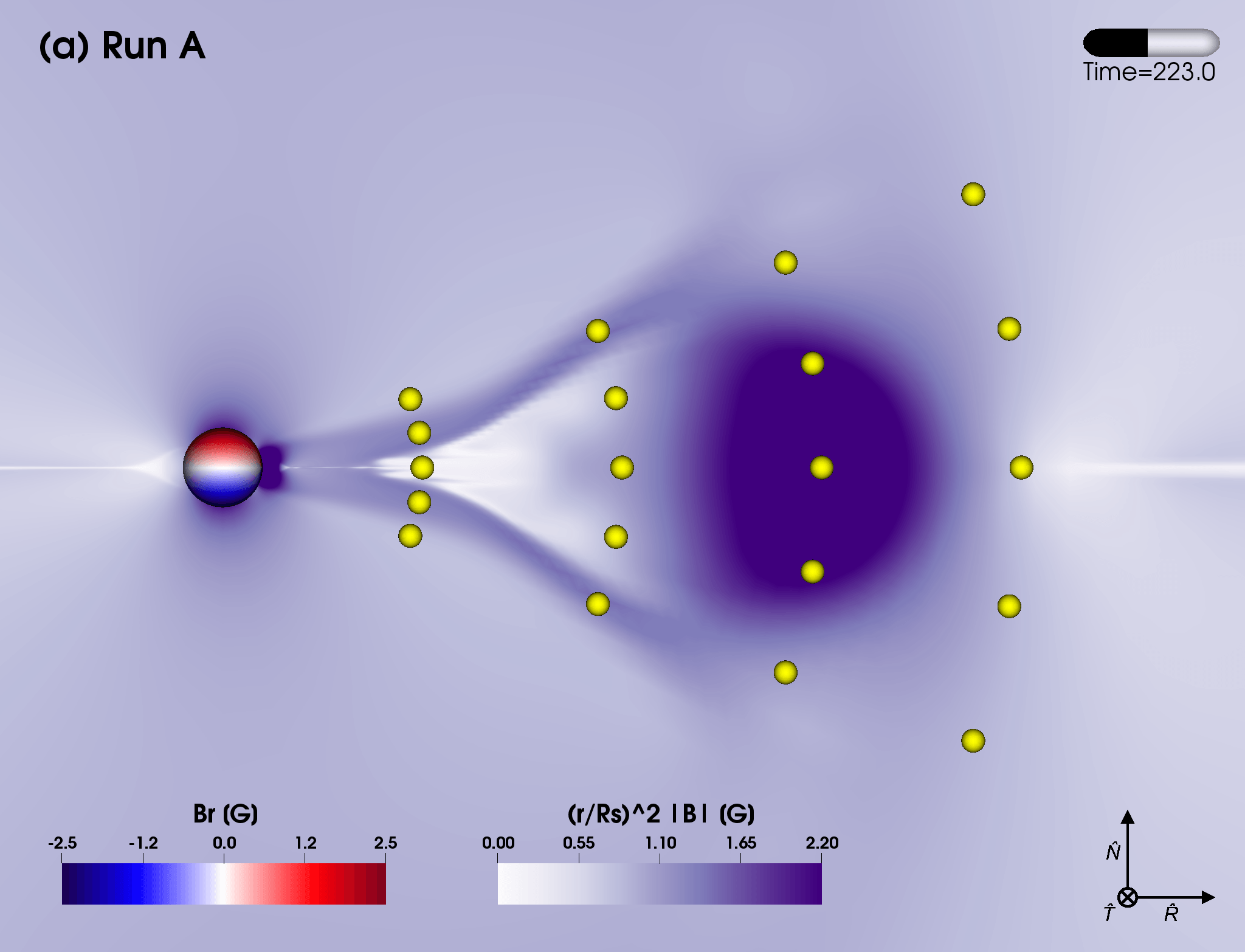}
\includegraphics[width=0.495\linewidth]{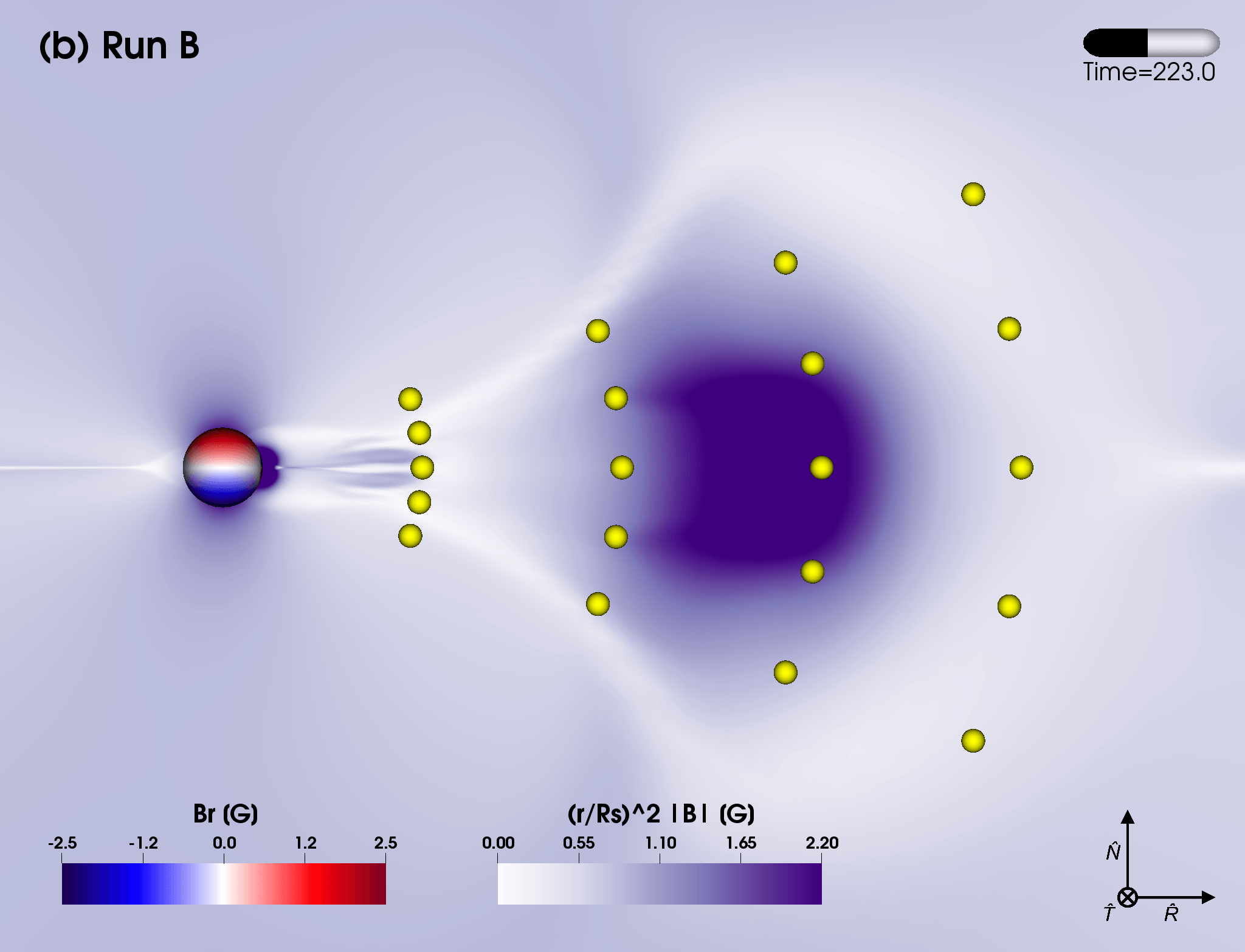}
\caption{Snapshot (at $t \simeq 9.5$~h) of the scaled magnetic field magnitude ($|B|$) for (a) Run~A and (b) Run~B projected onto the meridional plane at $\phi = 0^{\circ}$. The synthetic spacecraft that cross the meridional plane are represented by yellow spheres. The animated version of this figure shows the evolution of $|B|$ throughout the temporal domain of the simulations (0--20~h).\\
(An animation of this figure is available.)
\label{fig:btotslice} }
\end{figure*}

To determine conclusively whether this opposite-polarity field is part of the flux rope at least at the time of the in-situ crossing, we examine the evolution of the magnetic field lines connected to the 20 meridional synthetic spacecraft, as shown in Figure~\ref{fig:fieldlines}. From both the figure and its animated version, it is evident that, at the time of the encounter, the relevant field lines exhibit a highly complex topology, characterised by multiple twistings and foldings in and out of the core---indicating that they are part of the propagating CME magnetic structure. This suggests that the opposite-polarity component does not simply arise from external draping of the ambient coronal fields (which would produce a relatively stable, smooth field rotation), but instead constitutes an intrinsic part of the evolving flux rope system (which is measured as a larger and more rapid field rotation). Such a configuration may result from reconnection-driven restructuring of the erupting magnetic bundle and/or from the pre-eruptive and eruption-induced return-current layers surrounding magnetic flux ropes in the corona. \edit1{This scenario is also consistent with the findings of \citet{manchester2014a}, who showed that eruption-driven reconnection with the upstream fields at the CME leading edge could generate coherent sheared/twisted field structures along the interaction boundary and in the ejecta's wake.} The presence of these complex outer magnetic layers at the interface of the flux rope ejecta and the ambient medium---corresponding to (quasi-)separatrix surfaces \citep[e.g.,][]{priest1995, demoulin1996a, titov2008}---further highlights the extent to which CME magnetic structures can depart from the idealised force-free and symmetry assumptions in analytical flux rope models. Such complexity may naturally contribute to the ambiguous or unreliable fitting results discussed above, especially for crossings away from the CME nose and central axis.

\subsection{Consequences of Initial Topology} \label{subsec:topology}

As mentioned in Section~\ref{sec:setup}, both simulations employ a right-handed flux rope, but in Run~A (Run~B) the CME erupts from a bipolar (quadrupolar) source region. Here, we aim to explore topological differences between Run~A and Run~B stemming from the different initial configuration of the pre-eruptive source regions, to investigate how the coronal magnetic environment affects CME evolution in terms of propagation and internal structure. From the inspection of in-situ profiles discussed in Section~\ref{sec:results}, two main differences were identified between the two simulations (both visible in Figure~\ref{fig:cmenose}): the generally lower magnetic field magnitude in the Run~B ejecta and the high-speed flow following the Run~A CME.

\begin{figure*}[th!]
\centering
\includegraphics[width=0.495\linewidth]{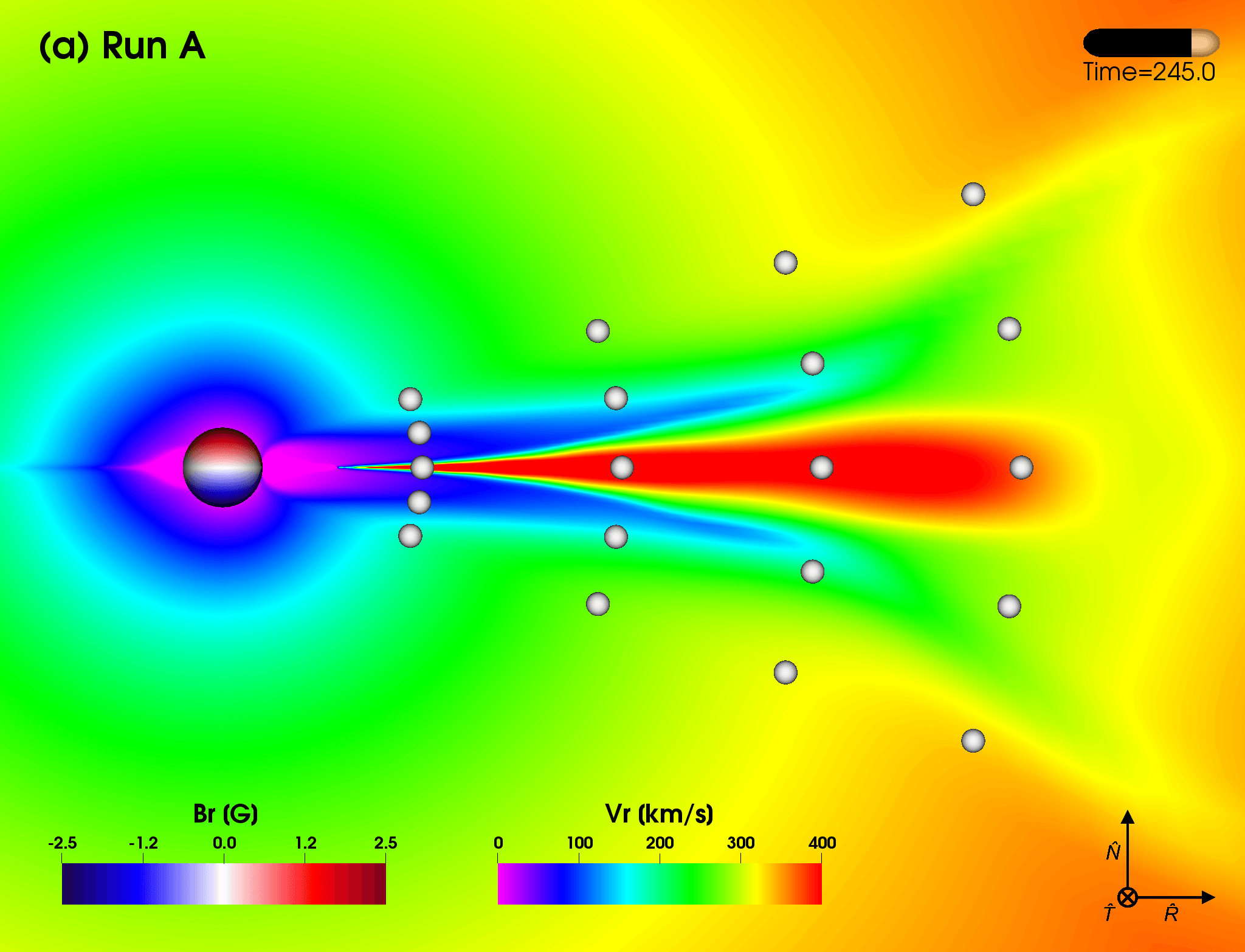}
\includegraphics[width=0.495\linewidth]{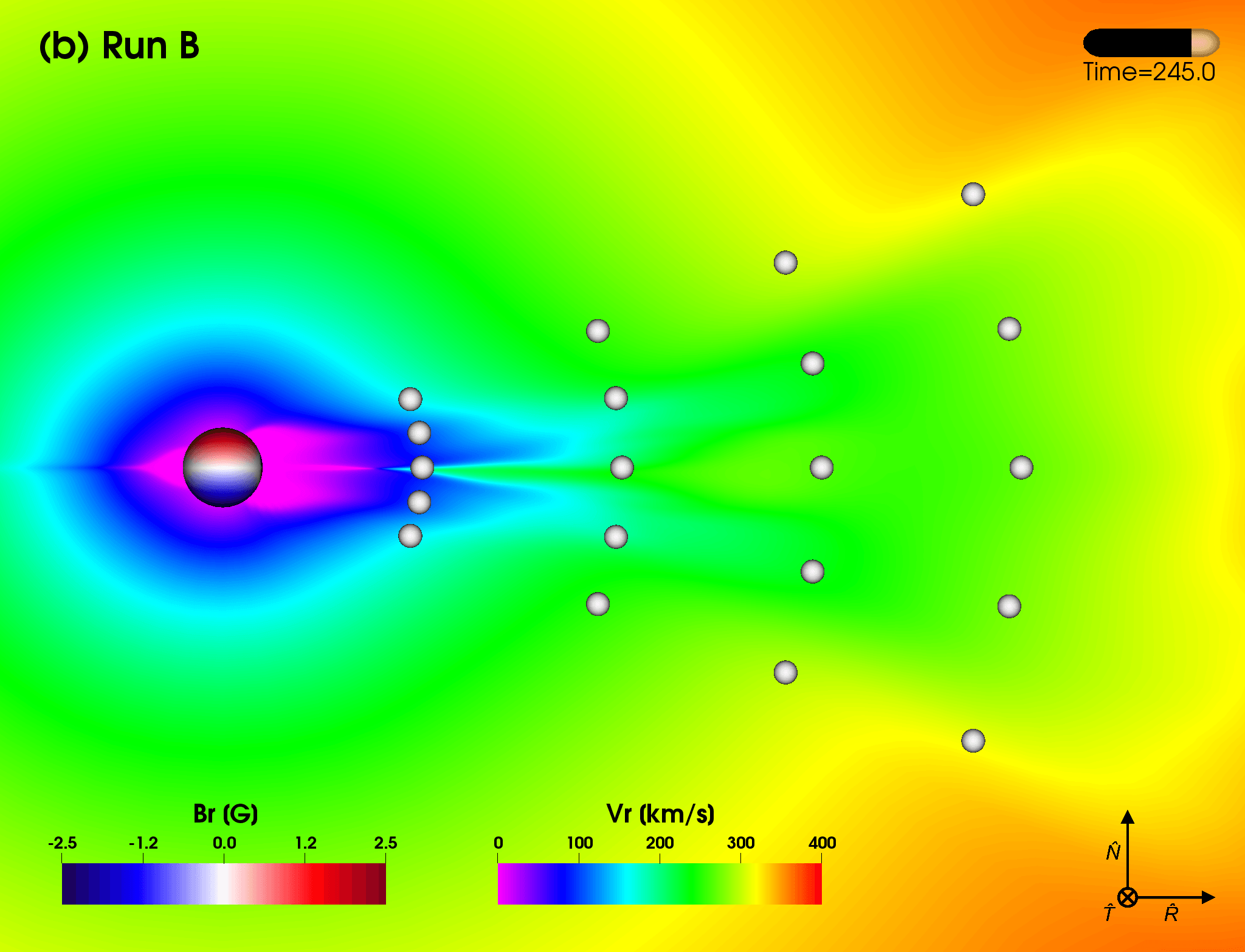}
\caption{Snapshot (at $t \simeq 18$~h) of the solar wind radial speed ($V_\mathrm{R}$) for (a) Run~A and (b) Run~B projected onto the meridional plane at $\phi = 0^{\circ}$. The synthetic spacecraft that cross the meridional plane are represented by white spheres. The animated version of this figure shows the evolution of $V_\mathrm{R}$ throughout the temporal domain of the simulations (0--20~h).\\
(An animation of this figure is available.)
\label{fig:vrslice} }
\end{figure*} 

To determine the origin of the different magnetic field strengths obtained for the two pre-eruptive configurations, we examine slices of the total magnetic field, shown for the meridional plane in Figure~\ref{fig:btotslice}. The figure and its accompanying animation reveal a notable difference between the two simulations. In Run~A, the magnetic structure exhibits the more familiar morphology of an erupting flux rope followed by a trailing current sheet. In Run~B, however, the CME is additionally surrounded by a broad region of weak magnetic field, indicative of ongoing reconnection in the enveloping current layers. This feature arises naturally from the quadrupolar configuration, in which the CME expands into a region of locally opposite magnetic polarity, leading to the formation of a strong current layer at the interface between the erupting flux rope and the ambient corona. The presence of this extended current sheet system has important consequences for the subsequent evolution of the CME magnetic structure. Returning to the field lines shown in Figure~\ref{fig:fieldlines}(b), it is apparent that field lines ahead of the eruption, initially belonging to the ambient open corona, reconnect with the leading edge of the CME as it propagates outward. By contrast, the field lines shown in Figure~\ref{fig:fieldlines}(a) show the CME ``plowing through'' the ambient corona rather than reconnecting with it, leading to substantially less erosion \citep[in agreement with][who found no significant erosion signatures for a simulated CME from a bipolar configuration]{manchester2014b}. In Run~B, favourable conditions for reconnection with the opposite-polarity ambient field lead to the progressive removal of magnetic flux from the front of the flux rope, resulting in erosion of the ejecta. \edit1{To quantify the degree of erosion that would be determined based on the in-situ signatures, we apply the ``direct method'' of \citet{dasso2006} to the central encounter profiles at 15\,$R_{\odot}$ (see, e.g., Figure~\ref{fig:latlonvar}) and calculate the estimated poloidal flux imbalance. We find that the Run~A and Run~B CMEs have undergone 8\% and 18\% erosion, respectively, consistent with the interpretation based on inspection of the magnetic field lines. Thus,} flux loss due to enhanced erosion offers a natural explanation for the systematically weaker magnetic field strengths observed in Run~B compared to Run~A and for the overall reduced size of the leading flux rope fields as is evident in Figure~\ref{fig:fieldlines}.

To identify the origin of the high-speed flow measured along the CME nose direction following the passage of the ejecta in Run~A, we examine slices of the radial velocity, shown for the equatorial plane in Figure~\ref{fig:vrslice}. The figure and its accompanying animation reveal a clear reconnection outflow trailing the flux rope in the bipolar configuration, whereas the corresponding feature in the quadrupolar case is much shorter-lived owing to the more rapid disconnection of the CME from its source region. This velocity enhancement is associated with the post-eruptive reconnection occurring beneath the erupting flux rope \citep[e.g.,][]{riley2007}. Such reconnection generates high-speed flows (and/or blobs) above the reforming loops of the flare arcade and, thus, velocity enhancements trailing the flux rope \citep[which can be observed even at 1~au; e.g.,][]{riley2002}. While this process is expected to accompany essentially all CME eruptions \citep{linker1995}, its duration and observational signature depend strongly on the magnetic configuration of the source region. In bipolar eruptions, the reconnection exhaust can remain magnetically connected to the propagating CME for an extended period, leading to an extended flare current sheet that undergoes tearing instability and results in a persistent (but bursty) high-speed flow trailing the ejecta \citep[e.g.,][]{murphy2013, jiang2021}. In quadrupolar configurations, by contrast, reconnection involving the side arcades contributes to the rapid reformation of the helmet streamer, leading to an earlier disconnection of the CME from the reconnection exhaust \citep[e.g.,][]{lynch2008, karpen2012}.

Finally, it is worth considering these findings in the context of a broader question: to what extent can the source region topology of a CME be inferred from its in-situ flux rope signature in the corona? Such a possibility is arguably most viable at coronal distances, where the erupting structure still retains a relatively direct connection to its solar source and evolves within an environment that preserves aspects of the large-scale photospheric magnetic configuration. At larger heliocentric distances, continued expansion, deformation, and interaction with the structured heliosphere progressively modify the CME magnetic content, making any unambiguous imprint of the source topology increasingly difficult to recover. The present results suggest that, in the coronal regime, secondary signatures associated with the eruption may provide additional information beyond the flux rope magnetic configuration itself. In particular, the two classes of behaviour identified in this study---the degree of magnetic flux erosion at the CME front and the presence of post-eruptive reconnection outflows---may carry information on the underlying magnetic configuration of the source region. Enhanced front-side erosion, facilitated by sustained reconnection between the CME and ambient open fields, is more readily produced in the quadrupolar configuration, where oppositely directed fields naturally favour prolonged interaction. Conversely, extended post-eruptive velocity enhancements in the wake of the ejecta, resulting from a longer-lived reconnection exhaust feeding outflows beneath the erupting flux rope, are more prominent in the bipolar case. Taken together, these signatures suggest a potential, albeit indirect, link between coronal source region topology and in-situ measurements: while the magnetic field structure of the flux rope alone may be insufficient to uniquely determine its origin, the combined evolution of magnetic flux and plasma flows provides additional constraints that may help discriminate between bipolar and multipolar eruption scenarios.


\section{Discussion and Contextualisation} \label{sec:discussion}

Although the simulations presented in this work employ idealised CME configurations embedded in a simplified coronal environment, they reproduce a number of features that are often observed in both numerical studies and spacecraft measurements. In particular, several characteristics identified in the synthetic in-situ profiles---including complex magnetic field rotations, multi-peaked magnetic signatures, and signatures associated with non-central flux-rope encounters---have also been reported in observations of real CMEs. This suggests that, despite their simplified nature, the simulations capture key aspects of CME magnetic structure and its in-situ manifestation. In this section, we discuss the broader implications of our findings and place them in the context of previous studies based on synthetic spacecraft measurements (Section~\ref{subsec:fake}) as well as observations of real CME events (Section~\ref{subsec:real}).

\subsection{Modelling Context} \label{subsec:fake}

A natural question arising from this work concerns the extent to which the synthetic profiles analysed here are representative of measurements that would be obtained by a real spacecraft. The profiles considered in this study were extracted from stationary observers in the corotating frame, and therefore neglect the relative motion of a spacecraft through the evolving CME structure. This is especially relevant for Parker Solar Probe, which, during its closest perihelion passages, reaches speeds of up to ${\sim}191$~km$\cdot$s$^{-1}$, substantially exceeding the local corotation speed (${\sim}20$~km$\cdot$s$^{-1}$ at $10\,R_{\odot}$). \citet{lynch2022} compared synthetic in-situ measurements obtained along realistic Parker Solar Probe trajectories in the corona with those extracted from stationary observers and found that the resulting magnetic field profiles remain qualitatively very similar, as the longitudinal displacement of the spacecraft during the encounter is generally small (${\sim}5^{\circ}$) compared to the angular extent of the CME (usually ${\sim}50^{\circ}$). \edit1{However, while these conclusions were drawn for encounters across a broad radial range (${\sim}$10--25\,$R_{\odot}$), the influence of the spacecraft trajectory may be significantly more pronounced near $10$\,$R_{\odot}$, where orbital speeds are roughly double those at ${\sim}20$\,$R_{\odot}$.} Applying several flux rope fitting techniques to 16 synthetic encounters spanning central and flank crossings of both low- and high-inclination flux ropes, \citet{lynch2022} reported broad agreement amongst fitting results for well-defined magnetic cloud-like profiles, while significantly larger variability emerged for more complex encounters. Their results suggest that analytical flux rope models provide a useful first-order description of CME magnetic structure in the corona, whilst also highlighting that the limitations and ambiguities associated with such approaches persist---and may become even more pronounced---below $30\,R_{\odot}$.

\begin{figure*}[th!]
\centering
\includegraphics[width=0.999\linewidth]{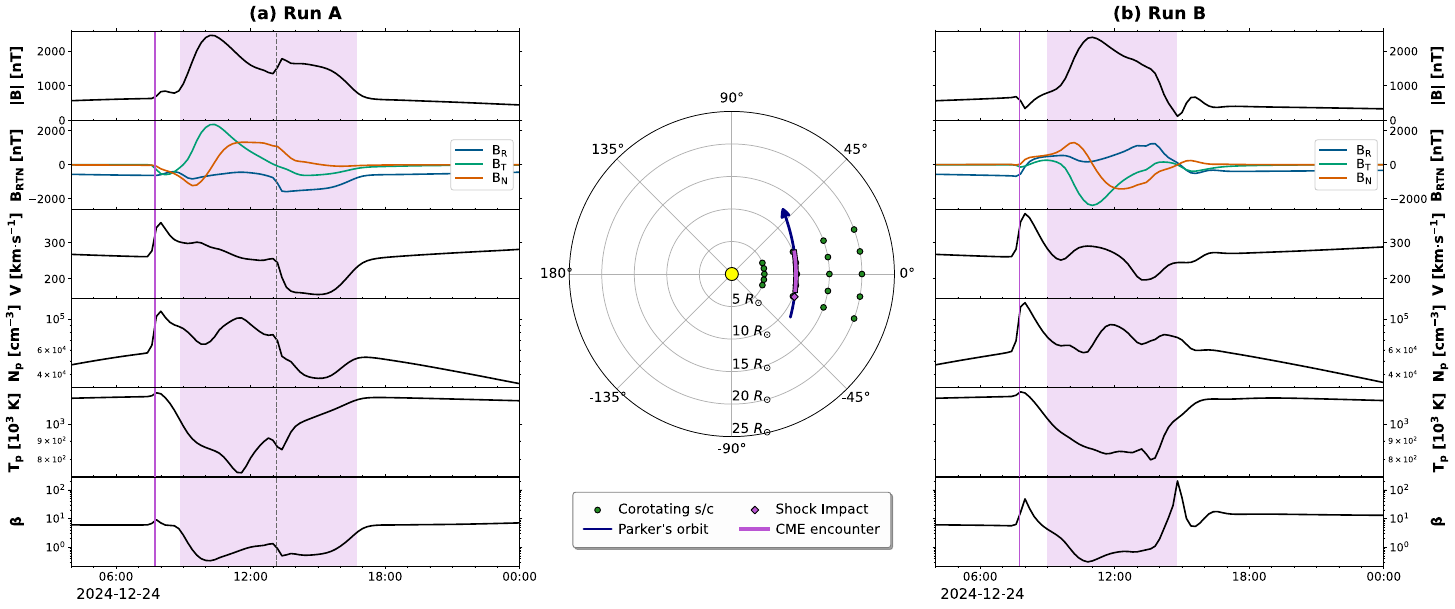}
\caption{Simulated Parker Solar Probe measurements of the (a) Run~A and (b) Run~B CMEs using the Encounter~\#22 trajectory. The in-situ profiles show, from top to bottom: the magnetic field magnitude, magnetic field components in RTN coordinates, solar wind speed, density, temperature, and plasma beta. CME-driven sheaths are marked by solid vertical lines, whilst CME flux rope intervals are indicated by shaded regions. The dashed vertical line in (a) highlights an ``ambiguous'' discontinuity region separating the two field magnitude enhancements. The central inset shows the Parker Solar Probe trajectory (in the Carrington frame) projected onto the equatorial plane, with the shock impact and CME passage (averaged between the two cases) times marked, together with the positions of the corotating synthetic spacecraft analysed throughout this work.
\label{fig:parker} }
\end{figure*} 

\edit1{To evaluate how our two events might appear during a realistic Parker Solar Probe encounter at ${\sim}10$\,$R_{\odot}$, we shift the trajectory of Encounter~\#22 (December 24, 2024)---the mission's first reach to its minimum perihelion---in longitude to achieve a near-central impact, as shown in Figure~\ref{fig:parker}. The resulting synthetic in-situ profiles appear somewhat more complex than the corotating profiles analysed throughout this work, blending characteristics from multiple stationary observers. In Run~A, the ejecta exhibits a double-peaked magnetic field magnitude---albeit less prominent than in the $\phi=\pm20^{\circ}$ cases of Figure~\ref{fig:lonvar}---which could be interpreted in real observations as either a complex encounter with a single structure or as a mixed region of CME and ambient material \citep[e.g.,][]{kay2026}, potentially leading to its exclusion from the CME ejecta analysis. In Run~B, low magnetic field regions appear at the ejecta boundaries, similar to stationary encounters with probes positioned further along the flanks. Additionally, both profiles show significantly more complex structuring in plasma parameters than in the nose encounters of Figure~\ref{fig:cmenose}. This complexity arises because Parker Solar Probe traverses essentially the entire longitude range sampled by our grid of stationary observers during the encounter (Figure~\ref{fig:parker}, central inset), whilst remaining at slightly southern latitudes throughout. Consequently, the non-stationary nature of the observer---unlike a spacecraft near 1~au, which can be approximated as quasi-stationary---yields mixed characteristics that complicate direct interpretation of the in-situ profiles. Nevertheless, the underlying flux rope type remains identifiable, with the magnetic field components showing clear SWN and NES rotations for Run~A and Run~B, respectively. Thus, the stationary profiles considered in this work can serve as guides to interpret localised portions of real Parker Solar Probe measurements, where the CME--spacecraft geometry can evolve rapidly \citep[e.g.,][]{mostl2020, braga2026}.}

Our findings can also be compared with those of \citet{rudisser2024}, who investigated the recovery of flux rope type from synthetic spacecraft measurements generated using an analytical CME model in the heliosphere. In contrast to the present study, they found that the intrinsic flux rope type could generally be recovered regardless of the spacecraft crossing location. This difference is likely a consequence of the underlying magnetic field description. In the analytical framework employed by \citet{rudisser2024}, the flux rope preserves a prescribed and well-defined magnetic structure throughout the analysis. In our simulations, by contrast, the erupting flux rope evolves self-consistently within a magnetised coronal environment, undergoing expansion, distortion, and magnetic reconnection with surrounding fields. As a result, the magnetic signatures sampled by synthetic spacecraft can differ substantially from those expected for an idealised flux rope. Despite these differences, \citet{rudisser2024} identified a number of flank encounters (for low-twist ropes only) for which the inferred flux rope type could be misclassified, including cases where a chirality opposite to that of the original structure may be recovered. Our results further support this finding and suggest that, in physics-based MHD simulations, these ambiguities may be further amplified as the CME magnetic structure is free to evolve and interact with the ambient corona.

Interestingly, a similar result was reported by \citet{riley2004}, who compared flux rope reconstructions performed independently by multiple groups for both a central and a flank encounter. While the central crossing yielded consistent results in terms of both chirality and axis orientation, substantially larger discrepancies emerged for the flank encounter, including one reconstruction that recovered the opposite handedness. At the time, this result was largely regarded as an outlier. Our simulations suggest that such discrepancies need not necessarily reflect deficiencies in the fitting procedure itself; rather, certain flank crossings may intrinsically contain insufficient or misleading information regarding the global magnetic topology of the CME, allowing even experienced analysts to reach fundamentally different interpretations.

\subsection{Observational Context} \label{subsec:real}

As mentioned at the beginning of this section, several of the synthetic signatures identified in this study have direct counterparts in spacecraft observations, suggesting that the magnetic structures and crossing geometries discussed in this work are not merely artefacts of the idealised setup. For example, both the ``double-crossing'' signatures associated with encounters spanning the CME front and leg, as well as the ambiguous flux rope classifications obtained for far-flank encounters, have been reported in real events.

We begin with crossings that remain relatively close to the flux rope axis but are displaced from the CME nose, such as the (${\theta}$, ${\phi}$) = ($0^{\circ}$, $-20^{\circ}$) profiles examined here (see Section~\ref{subsec:chirality} and Figure~\ref{fig:frfits}). A remarkably similar magnetic field signature was recently reported by \citet{palmerio2025} for a CME observed by BepiColombo at ${\sim}0.4$~au. In that event, the double-peaked $|B|$ profile consisted of an initial interval displaying a coherent flux-rope-like magnetic field rotation, followed by a second enhancement lacking a comparable smooth rotation, with the magnetic field components largely preserving their sign throughout the interval. Based on a combined analysis of remote-sensing and in-situ observations, \citet{palmerio2025} interpreted the event as a crossing of both the CME front and one of its legs. This interpretation is fully consistent with the geometry inferred from our simulations, where the double-peaked profiles arise naturally from trajectories that intersect the frontal portion of the flux rope before entering a leg region characterised by strong radial magnetic fields. Complex ``double-rotation'' magnetic field profiles were also reported in large statistical studies of CME measured at 1~au \citep[e.g.,][]{nieveschinchilla2019}, and they are usually interpreted as either interacting CMEs or complex single-CME configurations.

A more intriguing result concerns encounters far from the CME axis, such as the (${\theta}$, ${\phi}$) = ($20^{\circ}$, $0^{\circ}$) profiles analysed in this work. As discussed in Section~\ref{subsec:chirality}, these crossings can yield ambiguous flux rope classifications and, in some cases, signatures suggestive of a chirality opposite to that of the underlying global structure. Such trajectories may therefore help explain long-standing discrepancies between helicity inferred from solar source region observations and that recovered from in-situ measurements. For example, \citet{chandra2010} reported that the magnetic cloud associated with the November~18, 2003 CME exhibited a handedness opposite to that expected from its source region. To reconcile this discrepancy, the authors identified a localised region of positive helicity embedded within an otherwise predominantly negative-helicity active region and suggested that the eruption originated from this smaller-scale structure. An alternative interpretation of the same event was later proposed by \citet{uralov2014}, who argued that substantial magnetic restructuring and reconnection during the early phases of the eruption could account for the observed mismatch. While such explanations may indeed apply to individual events, our results demonstrate that apparent helicity reversals can also arise from the geometry of the spacecraft encounter itself, particularly when sampling the flanks of an evolving CME, where the local magnetic field configuration may differ substantially from that of the global flux rope structure.

Further observational support for this interpretation can be found in several more recent events. The March~15, 2013 CME analysed by \citet{pal2017} displayed a clear left-handed configuration at the Sun but right-handed flux rope characteristics at Earth. The pronounced asymmetry of the CME in coronagraph observations, together with the reconstruction of a highly inclined flux rope propagating predominantly eastward \citep[see Figure~5 of][]{pal2017}, suggests that the spacecraft may have sampled the structure away from its central axis in a manner analogous to the ${\theta}={\pm}20^{\circ}$ encounters examined here. Likewise, \citet{rodriguezgarcia2022} found that the August~19, 2013 CME appeared right-handed at the Sun and remained as such when observed by MESSENGER at 0.3~au and by STEREO-A at 1~au, yet was reconstructed as left-handed by STEREO-B, which was separated from STEREO-A by ${\sim}80^{\circ}$ in heliolongitude and therefore sampled a substantially different portion of the ejecta---interpreted in their study as a flank. Taken together, these observations support the notion that flank encounters can yield magnetic signatures that differ significantly from those expected on the basis of the global CME configuration alone.

In terms of signatures that may provide additional information on source region topology, an interesting example is the September 5, 2022 CME encountered by Parker Solar Probe at ${\sim}14\,R_{\odot}$. \citet{patel2025} reported a long-lived reconnection exhaust within the post-CME current sheet, indicating sustained, fast magnetic reconnection behind the erupting flux rope. Although the eruption occurred on the far side of the Sun as viewed from Earth, it was observed on the solar disc by Solar Orbiter \citep{long2023}, revealing that it originated from a bipolar active region, qualitatively consistent with the behaviour found in our bipolar simulation, where prolonged magnetic connectivity to the reconnecting current sheet produced an extended trailing high-speed flow. While a single event cannot establish a general relationship, it illustrates how coronal in-situ signatures associated with post-eruptive reconnection may offer useful clues regarding the magnetic topology of the source region. If such a connection proves robust in a larger sample of events, it may have implications beyond CME morphology alone, as prolonged magnetic connectivity to the reconnecting current sheet could influence plasma heating, the thermal structure of the ejecta, and the duration of energetic particle acceleration associated with the eruption \citep[e.g.,][]{lynch2011, reeves2019}.

\edit1{While flux rope erosion has been extensively quantified near 1~au \citep[e.g.,][]{ruffenach2015, pal2020}, estimates closer to the Sun remain scarce. One of the closest in-situ assessments was conducted by \citet{pal2022} for a CME encountered by Parker Solar Probe at ${\sim}0.5$~au, finding an $18\pm11$\% poloidal flux loss that initiated only beyond ${\sim}0.35$~au. Analysing an event close to 1~au, \citet{lavraud2014} inferred that up to ${\sim}50$\% of total magnetic erosion can occur beyond Mercury's orbit, further suggesting that interplanetary reconnection progressively erases signatures of the flux rope's original coronal environment. Further in-situ analyses of real CME encounters close to the Sun are thus essential to establish how source region topology may relate to reconnection rates in the corona.}


\section{Summary and Conclusions} \label{sec:conclusions}

In this work, we have presented a detailed analysis of the 3D magnetic structure of CMEs throughout the solar corona from an in-situ perspective. By producing synthetic spacecraft encounters with two different modelled eruptions at varying distances from the CME nose and axis, we have emulated different crossing geometries in order to assess how the inferred magnetic structure depends on the observer's location. The two simulated CMEs, previously presented by \citet{bennun2023}, represent flux ropes erupting from two different source configurations, i.e.\ a bipolar and a quadrupolar case. To isolate intrinsic CME properties from effects arising from interactions with a complex ambient environment, both simulations were embedded within a highly idealised solar minimum corona consisting of a dipolar Sun and a single active region located beneath the streamer belt. Our findings can be summarised as follows.

In the absence of complex solar wind structures such as in the configuration considered here, CMEs tend to evolve in a largely self-similar manner throughout the corona. Even in such a ``simple'' case, the retrieved in-situ profiles can vary substantially depending on the spacecraft crossing location, with more successful reconstructions of the CME structure obtained for encounters closer to the flux rope nose. Interestingly, we find cases in which the in-situ profiles could be misinterpreted as CME--CME interaction events (for crossings near the axis but far from the nose, effectively corresponding to an encounter through the front followed by one leg) or as flux ropes of opposite helicity to the true configuration (for flank encounters away from both the axis and the nose). This highlights that apparent complexity in in-situ CME signatures can arise purely from geometrical sampling effects rather than intrinsic changes in the magnetic structure of the ejecta, which remains, at the nose, largely consistent with the pre-eruptive configuration.

We also find clear differences between bipolar and quadrupolar CMEs in their in-situ signatures. Relative to the bipolar case, the quadrupolar eruption exhibits systematically weaker magnetic field strengths due to an additional enveloping current sheet, leading to enhanced flux erosion driven by sustained interaction and reconnection with the surrounding coronal magnetic field. In contrast, the bipolar CME is followed by a pronounced high-speed flow, clearly associated with a longer-lived post-eruptive reconnection exhaust that remains magnetically connected to the erupting structure. These differences are accompanied by more coherent and consistently recoverable flux rope signatures in the bipolar case, whereas the quadrupolar configuration exhibits more distorted magnetic profiles and larger fitting uncertainties. We suggest that identifying such signatures in coronal in-situ measurements may provide valuable diagnostic information on the magnetic topology of the CME source region.

These insights are particularly useful for interpreting CME encounters by Parker Solar Probe during its passages below ${\sim}30\,R_{\odot}$. The range of in-situ profiles explored here can be used to better constrain spacecraft crossing geometries in events measured in the corona and to support the interpretation of ambiguous signatures. This is especially relevant for far-sided eruptions, where the source region is not directly observable by solar disc imagers, yet knowledge of its magnetic configuration would provide important context for understanding the CME's evolution and associated processes such as particle acceleration \citep[e.g.,][]{mierla2022, dresing2025}. While real CME events encountered by Parker Solar Probe are expected to appear considerably more complex than the idealised cases examined here \edit1{(as is evident already from the synthetic profiles shown in Figure~\ref{fig:parker})}, we hope that the insights gained from this study will help distinguish intrinsic properties of the ejecta from signatures arising through interaction with a more structured ambient environment. More broadly, these results indicate that coronal in-situ observations can in principle retain information not only about the local magnetic structure sampled by the spacecraft, but also about the global geometry and evolutionary history of the eruption, providing a valuable bridge between remote-sensing observations and in-situ measurements at least close to the Sun.


\section*{Acknowledgments}
This work was supported by NSF's SHINE (grant no.\ AGS-2301403) programme as well as NASA's HSR (grant no.\ 80NSSC25M7101), LWS (grant no.\ 80NSSC24K1108), and LWS-SC (grant no.\ 80NSSC22K0893) programmes.

\vspace{5mm}


\bibliography{bibliography}{}
\bibliographystyle{aasjournal}

\end{document}